\UseRawInputEncoding

\documentclass[sn-mathphys]{sn-jnl}

\usepackage{epsfig}
\usepackage{graphicx}
\usepackage{hyperref}
\hypersetup{colorlinks=true,linkcolor=blue,citecolor=red,anchorcolor=black}
\usepackage{amsmath}
\usepackage{amssymb}
\usepackage{natbib}
\numberwithin{equation}{section}

\jyear{2022}

\begin{document}

\title{How to quantize gravity and how not to quantize gravity}

\author{\fnm{Philip D.} \sur{Mannheim}}\email{philip.mannheim@uconn.edu}
\affil{\orgdiv{Department of Physics}, \orgname{University of Connecticut},
\orgaddress{ \city{Storrs},  \state{Connecticut}, \postcode{06269}, \country{USA}}}

\abstract{
Taking the quantization of electromagnetism as the paradigm, we show how this procedure cannot work for Einstein gravity. However, it does
work for conformal gravity, a fourth-order derivative, renormalizable theory of gravity that Bender and Mannheim have shown to be ghost free. We show that in any gravity theory gravity cannot be quantized canonically. Rather, because of an interplay between the zero-point energy of gravity and that of its matter source, gravity is quantized purely by its coupling to a quantized matter source, with gravity being intrinsically quantum mechanical. Treating the zero-point energy this way provides a solution to the cosmological constant problem. With the gravitational zero-point energy issue having been ignored in standard Einstein gravity, it is not possible to solve the cosmological constant problem in standard gravity, since without the zero-point contribution gravity does not know where the zero of energy is.}

\maketitle

\section{\bf Quantum electrodynamics as the paradigm}

The Maxwell equations are really quite remarkable. They have remained intact following special relativity, following General Relativity and following quantum field theory. Originally written as 
\begin{eqnarray}
\boldsymbol{\nabla}\cdot\boldsymbol{E}=\rho,\quad \boldsymbol{\nabla}\times\boldsymbol{B}-\frac{\partial \boldsymbol{E}}{\partial t}=\boldsymbol{J},\quad
\boldsymbol{\nabla}\cdot\boldsymbol{B}=0,\quad \boldsymbol{\nabla}\times\boldsymbol{E}+\frac{\partial \boldsymbol{B}}{\partial t}=0,
\label{1.1}
\end{eqnarray}
they can be written in the special relativistic form 
\begin{eqnarray}
\partial_{\mu}F^{\mu\nu}=&-J^{\nu},\quad \partial_{\alpha}F_{\beta\gamma}+\partial_{\gamma}F_{\alpha\beta}+\partial_{\beta}F_{\gamma\alpha}=&0
\label{1.2}
\end{eqnarray}
as is. With General Relativity they are rewritten as 
\begin{eqnarray}
\nabla_{\mu}F^{\mu\nu}&=&-J^{\nu},
\nonumber\\
\nabla_{\alpha}F_{\beta\gamma}+\nabla_{\gamma}F_{\alpha\beta}+\nabla_{\beta}F_{\gamma\alpha}
&=&\partial_{\alpha}F_{\beta\gamma}+\partial_{\gamma}F_{\alpha\beta}+\partial_{\beta}F_{\gamma\alpha}=0.
\label{1.3}
\end{eqnarray}
And they do not get modified in form following quantization. What happens instead is that the fields are reinterpreted as q-number operators with nontrivial, $\hbar$-dependent, commutators.

The fact that the Maxwell equations  do not get modified in form is because quantum electrodynamics is a renormalizable field theory, with matrix elements of the quantum operators in states with an indefinite number of photons obeying the classical Maxwell field equations, albeit with renormalized parameters. This limit is not the limit in which we set $\hbar$ to zero, as $\hbar$ never is zero  in the real world.  
It instead is a many-particle macroscopic limit in which interference effects cause $\hbar$-dependent terms to mutually cancel each other, with the center of a wave packet for instance following the classical trajectory (Ehrenfest's theorem). Similarly, because of interference, a stationary phase approximation  to the Feynman path integral is just one path, the stationary classical path solution to the classical equations. 
While this stationary solution  itself does not depend on $\hbar$, the value of the path integral itself in the stationary solution, viz. $e^{iI_{STAT}/\hbar}$, does, where $I_{STAT}$ is the stationary classical action. As we will see below, this quantum-mechanical phase can be measured by gravity, the Colella-Overhauser-Werner experiment.

\subsection{\bf Classical electrodynamics follows from quantum electrodynamics and not vice versa}

\begin{figure}[h]
\centering
\includegraphics[scale=0.5]{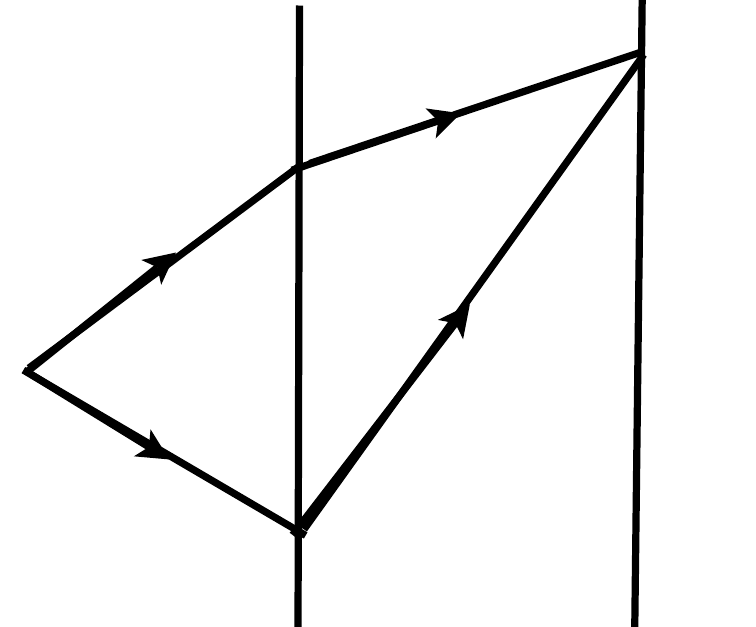}
\caption{Double slit experiment}
\label{slit}
\end{figure}

There are no separate classical and quantum theories of electrodynamics. Rather, classical electrodynamics is output to an underlying quantum electrodynamic theory. For instance, in the double slit experiment (Fig. \ref{slit}) the observed classical pattern on a screen is formed by an underlying quantum-mechanical behavior at the level of each individual  photon, with only their cumulative statistical distribution being observed because large numbers of photons arrive at the screen simultaneously. Each individual photon lands at one point on a screen, with a sequence of many photons then forming a pattern of maxima and minima on the screen, with the positions of the maxima and minima on the screen being given by the classical optical path differences.

Historically, quantum theory was developed by replacing classical  Poisson brackets by quantum commutators (canonical quantization). 
However, this is doing things backwards, classical follows from quantum not the other way round. 
Thus the task of quantizing gravity is to find a sensible microscopic quantum theory, with classical gravity being its many-graviton macroscopic limit. 
If that limit is to have the same form as the quantum gravitational equations, then the theory would need to be renormalizable (cf. conformal gravity), and  as has been observed by the present author and by John Donoghue, that would have been the natural choice had quantum field theory and $SU(3)\times SU(2) \times U(1)$ been developed prior to General Relativity.

We thus propose  to try to quantize gravity by replicating the treatment given for quantum electrodynamics. While this will work there will be a surprise: gravity cannot be quantized on its own.
Rather, it is quantized by its coupling to quantized matter, with there being no $\hbar^0$ term, and we should expand as a power series in Planck's constant and not as a power series in the gravitational coupling constant. 
This will depend heavily on making sense of the zero-point energy, and is central to solving the cosmological constant problem.

In terms of  path integrals,  for a gravitational field $g_{\mu\nu}$ and generic matter field $\psi$ we look for a gravitational path integral of the form 
\begin{eqnarray}
{\rm Gravitational~Path~Integral}=\int D[g_{\mu\nu}]D[\psi]e^{i[I_{\rm GRAV}+I_{\rm M}]/\hbar}
\label{1.4}
\end{eqnarray}
with classical gravitational and classical matter actions $I_{\rm GRAV}$ and $I_{\rm M}$.
We require that, for some appropriate choice of action and some appropriate domain for the path integral measure, the path integral be well defined to all orders of perturbation theory. And if we can find an appropriate action and an appropriate domain then we have a consistent theory of quantum gravity. 
This procedure does not work for Einstein gravity, and so one has to apply ``radical surgery" to quantum field theory and pursue a different approach such as string theory or loop quantum gravity.
However, it does work for conformal gravity, as it is renormalizable (it has a dimensionless coupling constant), and its ghost/unitarity problem is resolved \cite{bender2008no,bender2008exactly} by continuing the theory  into the complex plane, just as is needed to make the conformal gravity path integral exist.

\section{Does gravity actually know about quantum mechanics?}
\label{S}
\subsection{Experimental considerations}

Before discussing how one might quantize gravity we need to discuss whether we need to. Macroscopically, there are two established sources of  gravity that are intrinsically quantum-mechanical: (i) the Pauli degeneracy pressure of white dwarf stars with  Chandrasekhar mass $M_{\rm CH}\sim (\hbar c/G)^{3/2}/m_p^2$, and 
(ii) the energy density $\rho=\pi^2k_{\rm B}^4T^4/15c^3\hbar^3$ and pressure $p=\rho/3$ of the cosmic microwave background  black-body radiation in cosmology.
\begin{figure}[h]
\centering
\includegraphics[scale=1.0]{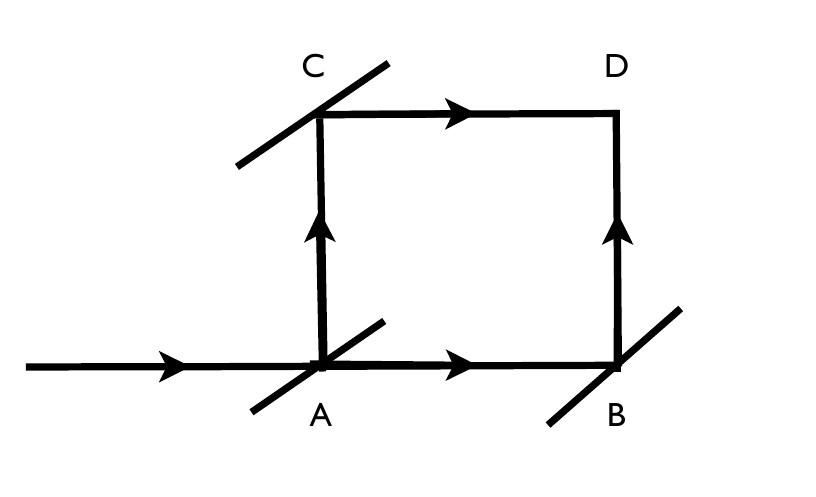}
\caption{Colella-Overhauser-Werner experiment}
\label{cowfull}
\end{figure}

Microscopically, the Colella-Overhauser-Werner experiment \cite{colella1975observation} shows that the quantum-mechanical phase of the wave function of a neutron of mass $m$ and velocity $v$ is modified as it traverses the gravitational field $g$ of the earth. In the vertical $ABCD$ interferometer shown in Fig. \ref{cowfull} $CD$ lies vertically above $AB$. The incoming neutron beam splits at $A$ with one component traveling  horizontally to $B$ and the other component traveling vertically to $C$. The components at $B$ and $C$ are then reflected so that they interfere at $D$. With $CD$ being at higher gravitational potential than $AB$, interference fringes are seen at $D$. With a change in the action of the neutron being of the form $\Delta I=-mgH^2/v$ (i.e., change compared to the $ABCD$ interferometer lying in the horizontal), the phase shift is given by  $\Delta \phi_{\rm COW}=\Delta I/\hbar=-mgH^2/v\hbar$ (see e.g. \cite{mannheim1998classical}), where $AB=BD=DC=CA=H$, yielding an  observable fringe shift at $D$ even though $H$ is only of the order of  centimeters and $m$ is the minuscule mass of a neutron --- it is just that $\Delta I$ is not small on the scale of Planck's constant.
 Thus gravity can measure the actual value of the stationary action and thus can measure the mass $m$, even though it drops out of the classical geodesic. The quantum-mechanical version of the equivalence principle is thus that the inertial and gravitational de Broglie wavelengths $\hbar/m_iv$ and $\hbar/m_gv$ are equal (i.e., interference in the horizontal and vertical). On measuring a nonvanishing fringe shift, Colella, Overhauser and Werner provided the first laboratory evidence of its kind that shows that gravity knows about quantum mechanics.

\subsection{Theoretical considerations and concerns}

In equations such as the Einstein equations: 
\begin{eqnarray}
-\frac{1}{8\pi G}\left(R^{\mu\nu} -\frac{1}{2}g^{\mu\nu}R^{\alpha}_{\phantom{\alpha}\alpha}\right)=T^{\mu\nu}_{\rm M},
\label{2.1}
\end{eqnarray}
we note that if the Einstein tensor is to be equal to the matter field energy-momentum tensor, then either both sides are classical c-numbers or both sides are quantum-mechanical q-numbers. Otherwise, if the gravity side were to be classical while the matter side were to be quantum mechanical, then the quantum $T^{\mu\nu}_{\rm M}$ would have to be equal to a c-number in every single field configuration imaginable. 
This is actually not impossible for a specific field configuration since a quantum-mechanical quantity such as the equal time field commutator $[\phi(t,\bar{x}),\pi(t,\bar{y})]=i\hbar Z\delta^3(\bar{x}-\bar{y})$ is a c-number -- it is just impossible if it has to be true for every field configuration. However, from gravitational experiments described above we know that the source of gravity is quantum-mechanical, and not only that, we know that gravity knows it. 
Hence gravity must be quantized.

However, since these gravitational experiments are not sensitive to quantum gravity effects themselves (graviton loops), for phenomenological purposes it is conventional to use a hybrid in which we keep gravity classical but take c-number matrix elements of its source, to give 
\begin{eqnarray}
-\frac{1}{8\pi G}\left(R^{\mu\nu} -\frac{1}{2}g^{\mu\nu}R^{\alpha}_{\phantom{\alpha}\alpha}\right)=\langle T^{\mu\nu}_{\rm M} \rangle.
\label{2.2}
\end{eqnarray}
But $\langle T^{\mu\nu}_{\rm M} \rangle$ involves products of fields at the same point, so it is not finite. Thus we additionally subtract off the infinite zero-point part and take the source to be the normal-ordered $\langle T^{\mu\nu}_{\rm M}\rangle_{\rm FIN}=\langle T^{\mu\nu}_{\rm M}\rangle-\langle T^{\mu\nu}_{\rm M}\rangle_{\rm DIV}$ instead, to thereby yield:
\begin{eqnarray}
-\frac{1}{8\pi G}\left(R^{\mu\nu} -\frac{1}{2}g^{\mu\nu}R^{\alpha}_{\phantom{\alpha}\alpha}\right)=\langle T^{\mu\nu}_{\rm M}\rangle_{\rm FIN}.
\label{2.3}
\end{eqnarray}
It is in this subtracted form that the standard applications of gravity are made. Thus in $\Sigma (a^{\dagger}a+1/2)\hbar \omega $ we keep the $\Sigma a^{\dagger}a\hbar \omega $ term but ignore the $\Sigma (1/2)\hbar \omega $ zero-point energy density term in $\langle T^{00}_{\rm M}\rangle$, precisely as is done in determining the Chandrasekhar mass or the black body contribution to cosmology. Also we  ignore the zero-point pressure in the spatial $\langle T^{ij}_{\rm M}\rangle$. 

As of today there is no known justification for using (\ref{2.3}), with this subtracted hybrid not having been derived from a fundamental theory or having been shown to be able to survive quantum gravitational corrections. Moreover, it is this very hybrid that is used in cosmology, and the cosmological constant problem is then the phenomenological need  to make $\langle T^{00}_{\rm M}\rangle_{\rm FIN}$ be small. Absent a first-principles derivation of 
\begin{eqnarray}
-\frac{1}{8\pi G}\left(R^{\mu\nu} -\frac{1}{2}g^{\mu\nu}R^{\alpha}_{\phantom{\alpha}\alpha}\right)=\langle T^{\mu\nu}_{\rm M}\rangle-\langle T^{\mu\nu}_{\rm M}\rangle_{\rm DIV}=\langle T^{\mu\nu}_{\rm M}\rangle_{\rm FIN},
\label{2.4}
\end{eqnarray}
this is not the right starting point for attacking the cosmological constant  problem.

There is no apparent reason why the zero point energy density of the matter sector should not gravitate. Moreover, while one only needs to consider energy differences in flat space physics, in gravity one has to consider energy itself, with the hallmark of Einstein gravity being that gravity couples to everything. Hence for gravity one cannot ignore zero-point contributions. And if we throw them away then gravity does not know where the zero of energy is -- and that is what creates the cosmological constant problem.

Moreover, even if one does start with the subtracted hybrid, then the order $G$ contribution to gravity is given by the flat spacetime  
$\langle \Omega \vert T^{\mu\nu}_{\rm M} \vert \Omega\rangle_{\rm FIN}$, with Lorentz invariance allowing a finite flat spacetime 
$\langle \Omega \vert T^{\mu\nu}_{\rm M} \vert \Omega\rangle_{\rm FIN}$ to be of the generic form $ -\Lambda g_{\mu\nu}$. 
Thus even if one ignores the matter sector zero-point energy contributions one still has a vacuum energy problem; with the standard strong, electromagnetic, and weak interactions typically then generating a huge such $\Lambda$. We thus recognize two types of vacuum problem, zero-point and $ -\Lambda g_{\mu\nu}$ problems. Moreover, if we take gravity to be quantum-mechanical  (as we must), it too will have divergent zero-point contributions.  

As has been shown in \cite{mannheim2017mass} and as we discuss below, rather than the divergent zero-point energy density in the gravity sector  being yet another vacuum energy problem, instead it is its interplay with the matter field zero-point contribution that actually leads to a solution to the cosmological constant problem. Thus the cosmological constant problem arises entirely due to ignoring how $\langle T^{\mu\nu}_{\rm M}\rangle_{\rm FIN}$ got to be finite in the first place, and then using  $R^{\mu\nu} -(1/2)g^{\mu\nu}R^{\alpha}_{\phantom{\alpha}\alpha} =-8\pi G\langle T^{\mu\nu}_{\rm M}\rangle_{\rm FIN}$ as the starting point.
However, in order to be able to discuss gravitational zero-point contributions, we need to quantize gravity. As we now show, such quantization cannot be done canonically, with gravity actually being quantized by the source that it is coupled to, rather than by being quantized on its own.

\section{\bf Canonical quantization of fermion fields}

For massless flat space fermion fields with matter action $I_{\rm M}=\int d^4x i\hbar\bar{\psi}\gamma^{\mu}\partial_{\mu}\psi$, stationary fields obey $\delta I_{\rm M}/\delta \bar{\psi}=i\hbar\gamma^{\mu}\partial_{\mu}\psi=0$. We introduce creation and annihilation operators according to
\begin{align}
\psi(x,t)=&\sum_{s}\int \frac{d^3k}{(2\pi)^{3/2}}\left[b(k,s)u(k,s)e^{-ik\cdot x}+d^{\dagger}(k,s)v(k,s)e^{ik \cdot x}\right],
\nonumber\\
\psi^{\dagger}(x,t)=&\sum_{s}\int \frac{d^3k}{(2\pi)^{3/2}}\left[b^{\dagger}(k,s)u^{\dagger}(k,s)e^{ik\cdot x}+d(k,s)v^{\dagger}(k,s)e^{-ik \cdot x}\right].
\label{3.1}
\end{align}
Quantizing the fermion field according to $\{\psi_{\alpha}(x,t),\psi_{\beta}^{\dagger}(x^{\prime},t)\}=\delta^3(x-x^{\prime})\delta_{\alpha,\beta}$ then requires that its creation and annihilation operators obey
\begin{eqnarray}
\{b(k,s),b^{\dagger}(k^{\prime},s^{\prime})\}&=&\delta^3(k-k^{\prime})\delta_{s,s^{\prime}},
\nonumber\\
 \{d(k,s),d^{\dagger}(k^{\prime},s^{\prime})\}&=&\delta^3(k-k^{\prime})\delta_{s,s^{\prime}}.
\label{3.2}
\end{eqnarray}

We construct the fermion field energy-momentum tensor by varying the same fermion action $I_{\rm M}$ (as momentarily written in curved space) with respect to a background metric $g_{\mu\nu}$. In the Fock space vacuum the vacuum matrix element of the energy-momentum tensor is given by
\begin{eqnarray}
\langle \Omega \vert \left[2g^{-1/2}\frac{\delta I_{\rm M}}{\delta g^{\mu\nu}}\right] \vert \Omega\rangle
&=&\langle \Omega\vert T^{\rm M}_{\mu\nu}\vert \Omega\rangle
\nonumber\\
&=&\langle \Omega\vert \bar{\psi}i\hbar \gamma_{\mu}\partial_{\nu}\psi \vert \Omega\rangle
=-\frac{2\hbar}{(2\pi)^3}\int \frac{d^3k}{k_0} k_{\mu}k_{\nu}.
\label{3.3}
\end{eqnarray}
We can identify zero point energy density and pressure terms of the form
\begin{eqnarray}
\rho_{\rm M}&=&\langle \Omega \vert  T^{\rm M}_{00}\vert \Omega \rangle =-\frac{2\hbar}{(2\pi)^3}\int d^3k \omega_k,
\nonumber\\
p_{\rm M}&=&\langle \Omega \vert  T^{\rm M}_{11}\vert \Omega \rangle =\langle \Omega \vert  T^{\rm M}_{22}\vert \Omega \rangle
\nonumber\\
&=&\langle \Omega \vert  T^{\rm M}_{33}\vert \Omega \rangle=-\frac{2\hbar}{(2\pi)^3}\int \frac{d^3k }{\omega_k}\frac{k^2}{3}=\frac{\rho_{\rm M}}{3}.
\label{3.4}
\end{eqnarray}
Thus we  have two zero-point infinities to deal with, viz.  $\rho_{\rm M}$ and $p_{\rm M}$, and neither behaves like a cosmological constant (for which $p=-\rho$), since $p_{\rm M}$ and $\rho_{\rm M}$ have the same sign. 

Now the fact that $\langle \Omega\vert T^{\rm M}_{\mu\nu}\vert \Omega\rangle$ is nonzero is not of concern in flat space field theory, and one can ignore it. Hence canonical quantization is based on stationarity with respect to the matter fields, and it is not of concern if it forces products of fields in the energy-momentum tensor to lead to a nonzero $\langle \Omega\vert T^{\rm M}_{\mu\nu}\vert \Omega\rangle$ since it can be normal-ordered away. But, what happens if gravity is present, and does this same canonical prescription work for gravity?

\section{\bf Quantization of the gravitational field}

\subsection{\bf Quantization of the gravitational field cannot be canonical}

For gravity let us introduce some general coordinate scalar action function of the metric $I_{\rm GRAV}$, and let us identify its variation
\begin{eqnarray}
\frac{2}{g^{1/2}}\frac{\delta I_{\rm GRAV}}{\delta g_{\mu\nu}}=T^{\mu\nu}_{\rm GRAV}
\label{4.1}
\end{eqnarray}
with respect to the metric as the energy-momentum tensor of gravity itself. 
(Such an identification has been suggested in  e.g. \cite{weinberg1972gravitation} and \cite{mannheim2006gauge}.) While the prescription is generic, for the Einstein-Hilbert action for instance we have 
\begin{eqnarray}
I_{\rm EH}=-\frac{1}{16 \pi G}\int d^4x (-g)^{1/2}R^{\alpha}_{\phantom{\alpha}\alpha},
\label{4.2}
\end{eqnarray}
\begin{eqnarray}
T^{\mu\nu}_{\rm GRAV}=\frac{1}{8 \pi G}\left[R^{\mu\nu}-\frac{1}{2}g^{\mu\nu}R^{\alpha}_{\phantom{\alpha}\alpha}\right].
\label{4.3}
\end{eqnarray}
For the general $I_{\rm GRAV}$ the gravitational equations of motion take the form:
\begin{eqnarray}
T^{\mu\nu}_{\rm GRAV}=0,
\label{4.4}
\end{eqnarray}
and in a typical linearization around flat spacetime of the form $g^{\mu\nu}=\eta^{\mu\nu}+h^{\mu\nu}$ would lead to a first-order wave equation of the form 
\begin{eqnarray}
T^{\mu\nu}_{\rm GRAV}(1)=0.
\label{4.5}
\end{eqnarray}
For the Einstein-Hilbert action for instance we would obtain $\nabla_{\alpha}\nabla^{\alpha} h_{\mu\nu}=0$ (in the harmonic gauge).

If we now were to quantize $h_{\mu\nu}$ canonically by identifying the coefficients of its negative and positive frequency modes as  creation and annihilation operators, we would find that the term that is second order in $h_{\mu\nu}$, viz. $T^{\mu\nu}_{\rm GRAV}(2)$, would not only not be zero, its vacuum expectation value would actually possess zero-point infinities. However, since the stationarity condition $T^{\mu\nu}_{\rm GRAV}=0$ is to hold to all orders in $h_{\mu\nu}$, stationarity with respect to the metric requires that 
\begin{eqnarray}
T^{\mu\nu}_{\rm GRAV}(2)=0.
\label{4.6}
\end{eqnarray}
Now for gravity we cannot normal order away its zero-point contributions since gravity couples to gravity, and thus we have a contradiction since canonical quantization yields
\begin{eqnarray}
T^{\mu\nu}_{\rm GRAV}(2)\neq 0.
\label{4.7}
\end{eqnarray}
We anyway should not get rid of $T^{\mu\nu}_{\rm GRAV}(2)$ since the one-graviton matrix element  of $T^{00}_{\rm GRAV}(2)$ is the energy carried by a gravity wave. 
Hence unlike all other fields for which the zero-point contribution does not conflict with canonical quantization, for gravity there is a conflict, since its stationarity condition is a statement about its zero point contribution, and this zero-point contribution involves an infinite product of fields at the same point. Hence, gravity cannot be quantized canonically, and in any quantization prescription for it one would have to take into account its zero-point infinities.

\subsection{Quantization of gravity via coupling to a quantized matter source}

Consider a generic action for the universe consisting of gravitational plus  matter actions of the form:
\begin{eqnarray}
I_{\rm UNIV}=I_{\rm GRAV}+I_{\rm M}.
\label{4.8}
\end{eqnarray}
Stationary variation with respect to the metric yields
\begin{eqnarray}
T^{\mu\nu}_{\rm UNIV}=T^{\mu\nu}_{\rm GRAV}+T^{\mu\nu}_{\rm M}=0.
\label{4.9}
\end{eqnarray}
If both $T^{\mu\nu}_{\rm GRAV}$ and $T^{\mu\nu}_{\rm M}$ are well-defined quantum-mechanically, we can treat the condition $T^{\mu\nu}_{\rm UNIV}=0$ as a quantum-mechanical operator identity, with its finite and divergent parts separately having to obey
\begin{eqnarray}
(T^{\mu\nu}_{\rm GRAV})_{\rm DIV}+(T^{\mu\nu}_{\rm M})_{\rm DIV}=0,
\label{4.10}
\end{eqnarray}
\begin{eqnarray}
(T^{\mu\nu}_{\rm GRAV})_{\rm FIN}+(T^{\mu\nu}_{\rm M})_{\rm FIN}=0.
\label{4.11}
\end{eqnarray}

Moreover, even if $T^{\mu\nu}_{\rm GRAV}$ is not well-defined quantum-mechanically, we can still apply $T^{\mu\nu}_{\rm GRAV}+T^{\mu\nu}_{\rm M}=0$ if we can (and below in fact do) make it be well defined quantum mechanically. Now for massless fermions  $\langle \Omega\vert T^{\mu\nu}_{\rm M}\vert \Omega \rangle$ is given by  the quartically divergent
\begin{eqnarray}
\langle \Omega\vert T^{\rm M}_{\mu\nu}\vert \Omega\rangle=-\frac{2\hbar}{(2\pi)^3}\int \frac{d^3k}{k_0} k_{\mu}k_{\nu},
\label{4.12}
\end{eqnarray}
\begin{eqnarray}
\rho_{\rm M}&=&\langle \Omega \vert  T^{\rm M}_{00}\vert \Omega \rangle =-\frac{2\hbar}{(2\pi)^3}\int d^3k \omega_k=-\frac{\hbar K^4}{4\pi^2},
\nonumber\\
p_{\rm M}&=&\langle \Omega \vert  T^{\rm M}_{11}\vert \Omega \rangle =-\frac{2\hbar}{(2\pi)^3}\int \frac{d^3k }{\omega_k}\frac{k^2}{3}=-\frac{\hbar K^4}{12\pi^2},
\label{4.13}
\end{eqnarray}
whee $K$ is a momentum cutoff.
Thus for four-component fermions we get a  zero-point energy of $-2\hbar \omega$ for each  momentum state $\bar{k}$, i.e. $-\hbar \omega $ for each of two two-component fermions. In order to cancel the $-2\hbar \omega$ fermion zero-point contribution we need to generate $+2\hbar \omega$ in the gravity sector $\langle \Omega\vert T^{\mu\nu}_{\rm GRAV}\vert \Omega \rangle$. Since the graviton zero-point contribution is quadratic in $h_{\mu\nu}$, we need $h_{\mu\nu}$ to be of order $\hbar^{1/2}$. 

Since the first-order equation of motion for $h_{\mu\nu}$, viz. $\langle \Omega\vert T^{\mu\nu}_{\rm GRAV}(1)\vert \Omega \rangle=0$ is homogeneous in $h_{\mu\nu}$, in and of itself it does not force us to have to choose a nontrivial solution to $\langle \Omega\vert T^{\mu\nu}_{\rm GRAV}(1)\vert \Omega \rangle=0$, and we can satisfy $\langle \Omega\vert T^{\mu\nu}_{\rm GRAV}(1)\vert \Omega \rangle=0$ with $h_{\mu\nu}=0$. 

Moreover, we would have to set $h_{\mu\nu}=0$ if we tried to quantize gravity on its own, since we would then need to also satisfy the second-order  $\langle \Omega\vert T^{\mu\nu}_{\rm GRAV}(2)\vert \Omega \rangle=0$. However, if gravity is coupled to a matter source, then the nonvanishing of the matter $\langle \Omega\vert T^{\mu\nu}_{\rm M}(2)\vert \Omega \rangle$ requires that the gravitational $\langle \Omega\vert T^{\mu\nu}_{\rm GRAV}(2)\vert \Omega \rangle$ be nonvanishing too. Hence coupling to a matter source forces gravity to have to choose the nontrivial solution to the first-order $\langle \Omega\vert T^{\mu\nu}_{\rm GRAV}(1)\vert \Omega \rangle=0$, with $h_{\mu\nu}$ then being forced to nontrivially be of order $\hbar^{1/2}$. Gravity is not quantized on its own. It is quantized by being coupled to a quantum-mechanical source.

To see how the quantization works in detail, for Einstein gravity for instance, we have one massless graviton with two helicity components, but a zero point of only $+\hbar\omega/2$ for each  momentum state, to give a total of $+\hbar\omega$. Thus we need to quantize the graviton field according to the generic  
\begin{eqnarray}
[\phi(t,\bar{x}),\pi(t,\bar{y})]=i\hbar Z\delta^3(\bar{x}-\bar{y}),
\label{4.14}
\end{eqnarray}
and thus need to set 
\begin{eqnarray}
Z=2,
\label{4.15}
\end{eqnarray}
and not the canonical $Z=1$. Moreover, if the source of gravity has $N$ two-component fermions (i.e., $N/2$ four-component fermions) and $M$ massless gauge bosons (with $+\hbar \omega /2$ for each of two helicity components), the net matter sector zero-point energy would be $-(N-M)\hbar \omega$, and so for the gravitational field we would have to set
\begin{eqnarray}
Z=N-M.
\label{4.16}
\end{eqnarray}
The normalization of canonical commutators  or anti-commutators in  the matter sector  and the matter content in the matter sector thus fixes the normalization in the gravity sector, with the gravity sector adjusting to whatever its source might be.  It is this interplay that makes $T^{\mu\nu}_{\rm GRAV}+T^{\mu\nu}_{\rm M}$ be well defined.

Still to be addressed is what is to happen when we take higher-order gravitational corrections into consideration, and what is to happen when we generate masses dynamically via a change in vacuum from a normal vacuum $\vert N\rangle$ to a spontaneously broken vacuum $\vert S \rangle$. 
To address these issues we need a theory of gravity in which radiative corrections are under control,  and so we turn now to conformal gravity, as it is a renormalizable theory of gravity, one for which mass generation would have to be spontaneous since unbroken conformal symmetry requires that all masses be zero. 
Such an unbroken symmetry also requires that any fundamental cosmological constant be zero, to thus provide a very good starting point to attack the cosmological constant problem.

Moreover, in regard to conformal invariance in general, the fermion-gauge boson sector of the standard $SU(3)\times SU(2)\times U(1)$ is conformal invariant at the level of the Lagrangian. In fact, this is the reason that they are second-order gauge theories in the first place, with conformal invariance then requiring dimensionless coupling constants and thus renormalizability. 
However, the double-well potential with its $-\mu^2\phi^2$ term would violate the conformal symmetry, and thus it must be excluded, with the symmetry breaking then being through dynamical fermion bilinears; to then (a la \cite{nambu1961dynamical}) yield dynamical fermion masses, dynamical massless Goldstone and massive Higgs bosons as fermion-antifermion bound states (a recent calculation in a four-fermion theory made renormalizable by anomalous dimensions may be found in \cite{mannheim2017mass}). 

Now the standard Einstein gravitational theory is not conformal invariant (which is why it is not renormalizable), and so we turn now to a gravity theory that is, viz. conformal gravity. Thus in the same way that $G_{\rm Fermi}$ is induced by dynamics we look to generate $G_{\rm Newton}$ dynamically too.

\section{Conformal gravity} 

Conformal gravity is based on the requirement of local conformal invariance of the action under $g_{\mu\nu}(x)\rightarrow e^{2\alpha(x)}g_{\mu\nu}(x)$. The Weyl conformal tensor  
\begin{eqnarray}
C_{\lambda\mu\nu\kappa}&=&R_{\lambda\mu\nu\kappa}
+\frac{1}{6}R^{\alpha}_{\phantom{\alpha}\alpha}\left[
g_{\lambda\nu}g_{\mu\kappa} -g_{\lambda\kappa}g_{\mu\nu}\right]
\nonumber\\
&-&\frac{1}{2}\left[
g_{\lambda\nu}R_{\mu\kappa} -g_{\lambda\kappa}R_{\mu\nu}
-g_{\mu\nu}R_{\lambda\kappa} +g_{\mu\kappa}R_{\lambda\nu}\right]
\label{5.1}
\end{eqnarray}
possesses a remarkable geometric property, namely that under a local conformal transformation on the metric it transforms  as $C_{\lambda\mu\nu\kappa} \rightarrow e^{2\alpha(x)}C_{\lambda\mu\nu\kappa}$ with all derivatives of $\alpha(x)$ dropping out identically. Given this fact, a locally conformal invariant  gravitational action of polynomial form can then uniquely be prescribed to be of the form:
\begin{eqnarray}
I_{\rm GRAV}&=&
-\alpha_g\int d^4x (-g)^{1/2}C_{\lambda\mu\nu\kappa} C^{\lambda\mu\nu\kappa}
\nonumber\\
&\equiv& -2\alpha_g\int d^4x (-g)^{1/2}\left[R^{\mu\nu}R_{\mu\nu}-\frac{1}{3}(R^{\alpha}_{\phantom{\alpha}\alpha})^2\right].
\label{5.2}
\end{eqnarray}
Then, since the coupling constant $\alpha_g$ is dimensionless, conformal gravity is power-counting renormalizable. 

Stationary variation of the conformal gravity action with respect to the metric leads to  gravitational equations of motion of the form 
\begin{eqnarray}
\frac{2}{(-g)^{1/2}}\frac{\delta I_{\rm GRAV}}{\delta g_{\mu\nu}}=T^{\mu\nu}_{\rm GRAV}=-4\alpha_g W^{\mu \nu}=-T^{\mu\nu}_{\rm M},
\label{5.3}
\end{eqnarray}
where (see e.g. \cite{mannheim1989exact,mannheim2006alternatives})
\begin{eqnarray}
W^{\mu \nu}&=&
\frac{1}{2}g^{\mu\nu}\nabla_{\beta}\nabla^{\beta}R^{\alpha}_{\phantom{\alpha}\alpha}
+\nabla_{\beta}\nabla^{\beta}R^{\mu\nu}                    
 -\nabla_{\beta}\nabla^{\nu}R^{\mu\beta}                       
-\nabla_{\beta}\nabla^{\mu}R^{\nu \beta}                          
 - 2R^{\mu\beta}R^{\nu}_{\phantom{\nu}\beta}                                    
\nonumber \\
&+&\frac{1}{2}g^{\mu\nu}R_{\alpha\beta}R^{\alpha\beta}-\frac{2}{3}g^{\mu\nu}\nabla_{\beta}\nabla^{\beta}R^{\alpha}_{\phantom{\alpha}\alpha}                                             
+\frac{2}{3}\nabla^{\nu}\nabla^{\mu}R^{\alpha}_{\phantom{\alpha}\alpha}                          
+\frac{2}{3} R^{\alpha}_{\phantom{\alpha}\alpha}R^{\mu\nu}   
\nonumber\\                           
&-&\frac{1}{6}g^{\mu\nu}(R^{\alpha}_{\phantom{\alpha}\alpha})^2,
\label{5.4}
\end{eqnarray}        
with both $W^{\mu\nu}$ and a conformal $T^{\mu\nu}_{\rm M}$ being both covariantly conserved and traceless.

A remarkable feature of (\ref{5.3}) is that when  $T^{\mu\nu}_{\rm M}$ is zero, (\ref{5.3}) admits of a solution in which $R_{\mu\nu}$ is zero. The conformal theory thus recovers the Ricci flat  Schwarzschild solution,  and thus meets the classical solar system tests of General Relativity without the use of the Einstein equations. However, the vanishing of the fourth-order derivative $W^{\mu\nu}$ allows for other solutions, with the general one having four integration constants and being of the form $-g_{00}=1/g_{rr}=w-u/r+vr-kr^2$ \cite{mannheim1989exact,mannheim2006alternatives}. The integration constants are not  dimensionless, and must be generated by spontaneous breakdown since the scale invariant $W^{\mu\nu}$ itself possesses no scale. In this way $G_{\rm Newton}$ is induced dynamically ($u=2MG_{\rm Newton}/c^2$). The linear and quadratic terms are important at large distance and enable the conformal theory to fit the galactic rotation curves of spiral galaxies with universal parameters and without either dark matter or its two free parameter per galaxy dark matter halos. In \cite{mannheim2011imp,mannheim2012fitgal,obrien2012fitdwa} very good fits to 138 galaxies were reported without the need for any of the 276 free parameters needed in dark matter fits to the same galaxy sample.

\section{Quantizing conformal gravity}
\subsection{Linearizing around flat}

If we linearize around flat spacetime with a transverse-traceless $K_{\mu\nu}$ we can set
\begin{eqnarray}
g^{\mu\nu}&=&\eta^{\mu\nu}+h^{\mu\nu},\qquad K^{\mu\nu}=h^{\mu\nu}-\frac{1}{4}\eta^{\mu\nu}\eta_{\alpha\beta}h^{\alpha\beta},\qquad \partial_{\mu}K^{\mu\nu}=0,
\nonumber\\
T^{\mu\nu}_{\rm GRAV}(1)&=&-4\alpha_gW^{\mu\nu}(1)=-2\alpha_g(\partial_{\alpha}\partial^{\alpha})^2 K^{\mu\nu}.
\label{6.1}
\end{eqnarray}
Setting $T^{\mu\nu}_{\rm GRAV}(1)=0$ has generic solution
\begin{eqnarray}
K^{\mu\nu}=A^{\mu\nu}e^{ik\cdot x}+(n\cdot x)B^{\mu\nu}e^{ik\cdot x},\qquad n^{\mu}=(1,0,0,0),
\label{6.2}
\end{eqnarray}
and full solution \cite{mannheim2011comprehensive}
\begin{eqnarray}
K_{\mu\nu}(x)&=&\frac{\hbar^{1/2}}{2(-\alpha_g)^{1/2}}\sum_{i=1}^2\int \frac{d^3k}{(2\pi)^{3/2}(\omega_k)^{3/2}}
\nonumber\\
&\times&\bigg{[}
A^{(i)}(\bar{k})\epsilon^{(i)}_{\mu\nu}(\bar{k})e^{ik\cdot x}
+i\omega_kB^{(i)}(\bar{k})\epsilon^{(i)}_{\mu\nu}(\bar{k})(n\cdot x)e^{ik\cdot x}
\nonumber\\
&+&\hat{A}^{(i)}(\bar{k})\epsilon^{(i)}_{\mu\nu}(\bar{k})e^{-ik\cdot x}
-i\omega_k\hat{B}^{(i)}(\bar{k})\epsilon^{(i)}_{\mu\nu}(\bar{k})(n\cdot x)e^{-ik\cdot x}\bigg{]},
\label{6.3}
\end{eqnarray}
where the fields are quantized according to
\begin{eqnarray}
&&[A^{(i)}(\bar{k}),\hat{B}^{(j)}(\bar{k}^{\prime})]=[B^{(i)}(\bar{k}),\hat{A}^{(j)}(\bar{k}^{\prime})]=Z(k)\delta_{i,j}\delta^3(\bar{k}-\bar{k}^{\prime}),
\nonumber\\
&&[A^{(i)}(\bar{k}),\hat{A}^{(j)}(\bar{k}^{\prime})]=0,\quad [B^{(i)}(\bar{k}),\hat{B}^{(j)}(\bar{k}^{\prime})]=0,
\nonumber\\
&& [A^{(i)}(\bar{k}),B^{(j)}(\bar{k}^{\prime})]=0,\quad [\hat{A}^{(i)}(\bar{k}),\hat{B}^{(j)}(\bar{k}^{\prime})]=0.
\label{6.4}
\end{eqnarray}
The fact that some of these commutators are zero is indicative of the Hamiltonian being of nondiagonalizable Jordan-block form with an incomplete set of eigenvectors, a point we return to below.

 Evaluating now the second-order term, we find that is associated with an action
\begin{eqnarray}
I_{\rm W}(2)=-\frac{\alpha_g}{2}\int d^4x \partial_{\alpha}\partial^{\alpha} K_{\mu\nu}\partial_{\beta}\partial^{\beta} K^{\mu\nu},
\label{6.5}
\end{eqnarray}
and yields 
\begin{eqnarray}
\langle \Omega \vert T_{\rm GRAV}^{\mu\nu}(2)\vert \Omega \rangle=-4\alpha_g\langle \Omega\vert W^{\mu\nu}(2)\vert \Omega \rangle=\frac{2\hbar}{(2\pi)^3} \int\frac{d^3k}{\omega_k}Z(k)k^{\mu}k^{\nu}.
\label{6.6}
\end{eqnarray}
Unlike Einstein gravity this time there are two graviton-type contributions, to thus give a zero-point of $+2\hbar \omega$. As previously
\begin{eqnarray}
T^{\mu\nu}_{\rm UNIV}=T^{\mu\nu}_{\rm GRAV}+T^{\mu\nu}_{\rm M}=0.
\label{6.7}
\end{eqnarray}
With one four-component massless fermion matter source, through order $\hbar$ we obtain
\begin{eqnarray}
\langle \Omega\vert T^{\mu\nu}_{\rm UNIV}\vert \Omega\rangle&=&\frac{2\hbar}{(2\pi)^3} \int \frac{d^3k}{\omega_k}Z(k)k^{\mu}k^{\nu}-
\frac{2\hbar}{(2\pi)^3}\int \frac{d^3k}{\omega_k}k^{\mu}k^{\nu}=0,
\nonumber\\
Z(k)&=&1.
\label{6.8}
\end{eqnarray}
As we see, $Z(k)$ is only nonzero because the matter zero point $\langle \Omega \vert T^{\mu\nu}_{\rm M}\vert \Omega\rangle$ is nonzero.

\subsection{Some phenomenology}

If we have $M$ massless gauge bosons and $N$ massless two-component spinors,  then gravity adjusts to its source to yield 
\begin{eqnarray}
Z(k)=\frac{(N-M)}{2},
\label{6.9}
\end{eqnarray}
with the positivity of $Z(k)$ entailing that $N>M$ (i.e., fermions are favored over gauge bosons).
For $SU(3)\times SU(2)\times U(1)$ we have M=12, N=16 per generation, so $N>M$ generation by generation. For SO(10) we have M=45, N=16, so we need 3 generations. For all generations in a common multiplet we need an $SO(2n)$ with $2n \geq 16$. (There is no solution for $SU(N)$ if all generations are in a single multiplet, so triangle-anomaly-free grand-unifying groups are preferred.)  For SO(16) we have M=120, N=128, so 8 generations. We can only have asymptotic freedom (which favors gauge bosons over fermions) up to $SO(20)$ if all fermions are to be in the same multiplet \cite{Wilczek1982families}. So quantum conformal gravity implies that the only possible grand-unifying groups with all fermions in the same multiplet are in the quite narrow window consisting of  SO(16), SO(18) and SO(20).

In general we have
\begin{eqnarray}
(T^{\mu\nu}_{\rm GRAV})_{\rm DIV}+(T^{\mu\nu}_{\rm M})_{\rm DIV}=0,
\label{6.10}
\end{eqnarray}
\begin{eqnarray}
(T^{\mu\nu}_{\rm GRAV})_{\rm FIN}+(T^{\mu\nu}_{\rm M})_{\rm FIN}=0.
\label{6.11}
\end{eqnarray}
Thus the zero-point fluctuations of gravity and matter take care of each other. Moreover, the divergent piece comes from the vacuum  while the finite piece comes from excitations out of the vacuum. Thus the measured cosmological constant in cosmology is just the finite, excitation out of the vacuum, piece, and not the entire finite plus divergent pieces combined.

Moreover, because of (\ref{6.9}), i.e., because conformal gravity adjusts to whatever its matter source is, trace anomalies take care of each other since the condition $g_{\mu\nu}T^{\mu\nu}_{\rm UNIV}=0$ is not a conformal Ward identity, but is instead an equation of motion, with the conformal gravity and matter sector trace anomalies having to cancel each other identically. Our ability to solve the trace-anomaly problem is because unlike in the standard case we do we not start with 
$(-1/8\pi G)G^{\mu\nu}=\langle T^{\mu\nu}_{\rm M}\rangle$ and try to then show that $\langle T^{\mu\nu}_{\rm M}\rangle$ is anomaly-free all on its own.

Since $T^{\mu\nu}_{\rm UNIV}=T^{\mu\nu}_{\rm GRAV}+T^{\mu\nu}_{\rm M}=0$ is an identity, it holds in any vacuum. Thus even after we spontaneously break the conformal symmetry, $(T^{\mu\nu}_{\rm GRAV})_{\rm FIN}$ will remain traceless since $W^{\mu\nu}$ is traceless, and thus so will $(T^{\mu\nu}_{\rm M})_{\rm FIN}$. This then constrains the cosmological constant that is induced by the change in vacuum, so that it is neither bigger nor smaller than any other contribution.

Now if all particle physics mass and length scales are dynamical, then all such scales are intrinsically quantum-mechanical. However, it is mass and length scales that characterize spacetime curvature. 
Hence in the absence of quantum mechanics there can be no curvature, and gravity must thus be intrinsically quantum-mechanical.

We thus need to expand the gravitational equations not as power series in the gravitational coupling constant but as a power series in Planck's constant, with no $\hbar^0$ term, just as we found with the effect of the zero point on quantization of the gravitational field.

\begin{figure}[h]
\epsfig{file=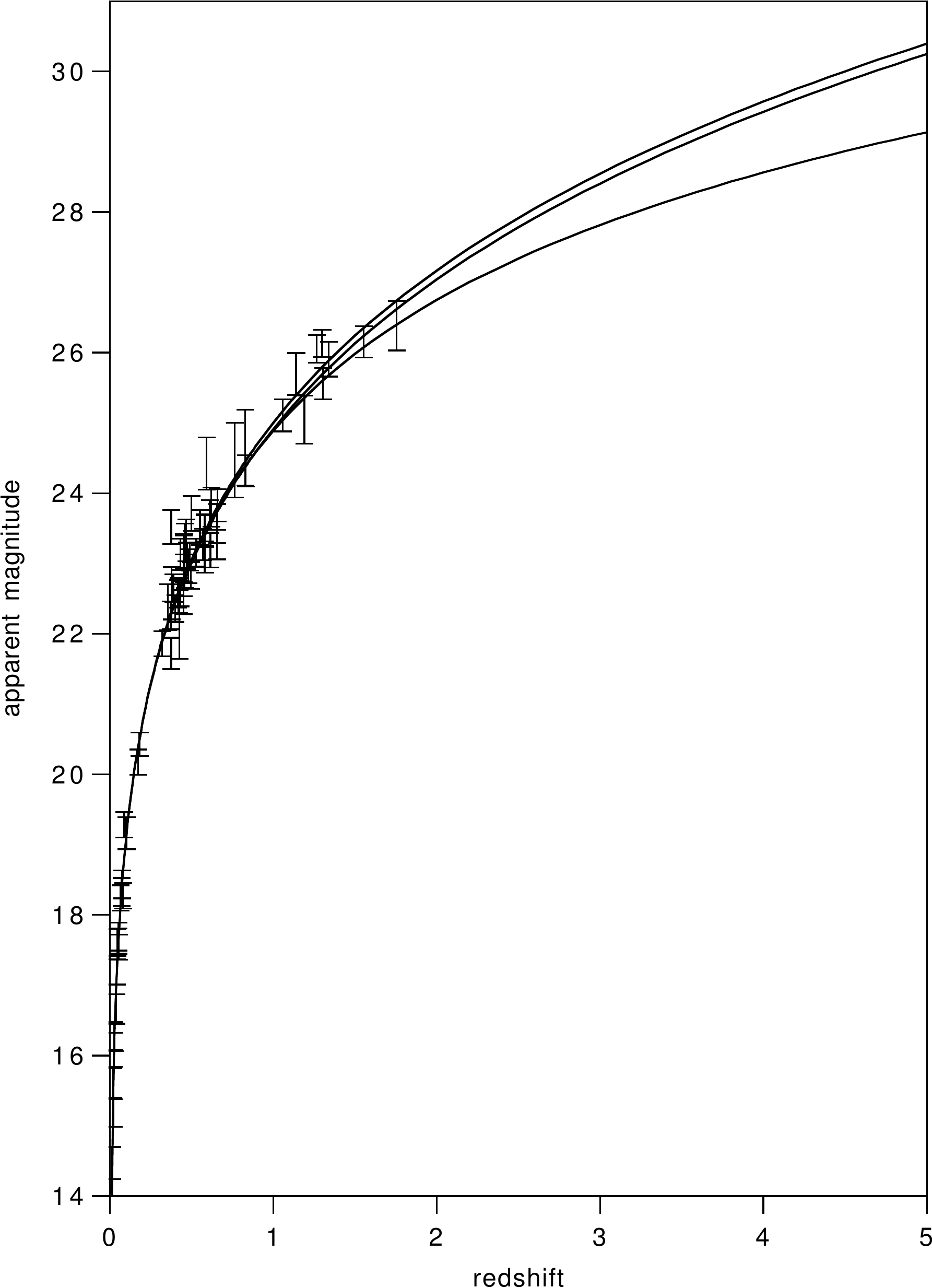,height=2.4in,width=4.0in}
\caption{Luminosity versus redshift Hubble plot expectations for $q_0=-0.37$ conformal gravity  (highest curve), for the empty universe
with $q_0=0$ (middle curve), and for
$\Omega_{M}(t_0)=0.3$,
$\Omega_{\Lambda}(t_0)=0.7$ standard gravity (lowest curve).}
\label{Hubbleplot}
\end{figure}

On solving the cosmology associated with conformal gravity in the presence of a cosmological constant  induced by the mass-generating spontaneous breakdown of the conformal symmetry, we obtain \cite{mannheim2006alternatives} a luminosity distance ($d_L$) versus redshift ($z$) relation of the form
\begin{eqnarray}
d_L= -\frac{c}{H_0}\frac{(1+z)^2}{q_0}\left(1-\left[1+q_0-\frac{q_0}{(1+z)^2}\right]^{1/2}\right),
\label{6.12}
\end{eqnarray}
with the current era deceleration parameter being $q_0$, and with a deceleration parameter $q$  that the conformal theory constrains to obey
\begin{eqnarray}
-1 \leq q \leq 0
\label{6.13}
\end{eqnarray}
in all epochs, no matter what values the parameters in $(T^{\mu\nu}_{\rm M})_{\rm FIN}$ take. I.e., the theory is naturally accelerating in all epochs without any fine-tuning. Equation (\ref{6.12}) is found \cite{mannheim2006alternatives} to give a very good fit (see Fig. \ref{Hubbleplot}) to the accelerating universe data \cite{Riess1998observational,Perlmutter1998measurements} without any fine tuning  of the cosmological constant; with the fitted value obtained for $q_0$ being found to be $q_0=-0.37$, viz. right in the $-1 \leq q_0 \leq 0$ range required by (\ref{6.13}). In the figure we have also included the standard $\Omega_{M}(t_0)=0.3$,
$\Omega_{\Lambda}(t_0)=0.7$ fit for comparison. As we see, at high enough redshift the two fits behave markedly differently. For further comparison purposes we have also included the so-called coasting or empty universe with $\Omega_{M}(t_0)=0$ and
$\Omega_{\Lambda}(t_0)=0$, a cosmology where $q_0=0$.

\section{Ghosts and unitarity}

Since the conformal gravity coupling constant $\alpha_g$ is dimensionless, conformal gravity is power-counting renormalizable. However, since the action contains four derivatives of the metric, the equations of motion are fourth-order derivative equations, and for a long time were thought to lead to states of negative norm (ghost states) and violations of unitarity. However, Bender and Mannheim \cite{bender2008no,bender2008exactly} have reanalyzed fourth-order theories and shown that they have to be reinterpreted as $PT$ theories in which the inner product is given as the overlap of a ket with its $PT$ conjugate rather than with its Hermitian conjugate; with the Hilbert space then being free of any states with either negative norm or  negative energy. Conformal gravity is thus recognized as a consistent quantum theory of gravity, one expressly formulated in four spacetime dimensions, the only spacetime  dimensions  for which we have any evidence.

\section{Why there cannot be a ghost problem}

Consider the Dirac action for a massless fermion coupled to a background geometry of the form
\begin{eqnarray}
I_{\rm D}=\int d^4x(-g)^{1/2}i\bar{\psi}\gamma^{c}V^{\mu}_c(\partial_{\mu}+\Gamma_{\mu})\psi, 
\label{8.1}
\end{eqnarray}
where the $V^{\mu}_a$ are vierbeins and $\Gamma_{\mu}=-(1/8)[\gamma_a,\gamma_b](V^b_{\nu}\partial_{\mu}V^{a\nu}+V^b_{\lambda}\Gamma^{\lambda}_{\nu\mu}V^{a\nu})$ is the spin connection that enables $I_{\rm D}$ to be locally Lorentz invariant. 

As constructed, $\Gamma_{\mu}$ also enables $I_{\rm D}$ to be  locally conformal invariant under 
\begin{align}
V^{\mu}_a\rightarrow e^{-\alpha(x)}V^{\mu}_a(x),~~~\psi(x)\rightarrow e^{-3\alpha(x)/2}\psi(x),~~~g_{\mu\nu}(x)\rightarrow e^{2\alpha(x)} g_{\mu\nu}(x).
\label{8.2}
\end{align}
We thus get local conformal invariance for free, with the massless fermion Dirac action naturally having an underlying local conformal symmetry. 't Hooft \cite{hooft2015local} has also argued that there should be an underlying local conformal symmetry in nature.

We now introduce the path integral $\int D[\psi]D[\bar{\psi}]\exp{iI_{\rm D}}=\exp(iI_{\rm EFF})$, and on performing the path integration on $\psi$ and $\bar{\psi}$ obtain an effective action with leading term  
\cite{'tHooft2010}
\begin{eqnarray}
I_{\rm EFF}&=&C\int d^4x(-g)^{1/2}\left[R_{\mu\nu}R^{\mu\nu}-\tfrac{1}{3}(R^{\alpha}_{\phantom{\alpha}\alpha})^2\right],
\label{8.3}
\end{eqnarray}
where $C$ is a log divergent constant. 

Thus in the standard model itself we generate the fourth-order derivative conformal gravity action. But the standard model is ghost free. Since the fermion path integral is equivalent to a one loop Feynman diagram and since one cannot change the signature of a Hilbert space in perturbation theory, conformal gravity must be ghost free too.
And if it were not the standard $SU(3)\times SU(2)\times U(1)$ model would not remain unitary when coupled to gravity.
Radiative corrections to $SU(3)\times SU(2)\times U(1)$ in an external gravitational field generate fourth-order conformal gravity whether we like it or not. So we have to deal with it one way or another.

\section{Why do people think that there is a ghost issue?}

We can write the conformal gravity $1/k^4$ propagator as the limit
\begin{eqnarray}
\frac{1}{k^4}=\lim_{M^2\rightarrow 0} \frac{1}{M^2}
\left(\frac{1}{k^2-M^2}-\frac{1}{k^2}\right),
\label{9.1}
\end{eqnarray}
with some of the poles having residues that are negative, to thus be associated with ghost states with negative norm. That there might be something wrong with this conventional reasoning can be seen by writing the $1/k^4$ propagator as a different limit
\begin{eqnarray}
\frac{1}{(k^2+i\epsilon)^2}=\lim_{M^2\rightarrow 0} \frac{d}{dM^2}
\left(\frac{1}{k^2-M^2+i\epsilon}\right),
\label{9.2}
\end{eqnarray}
and now there are no negative residue concerns. The first case has two sets of poles (i.e., two gravitons), while the second one only has one set of poles  (i.e., one graviton). So which is right?

\section{So what is wrong with the conventional reasoning?}

A propagator is a c-number and we cannot determine the structure of the underlying quantum theory from it. One might think that the propagator represents a matrix element such as $i\langle \Omega \vert T[\phi(x)\phi(0)]\vert \Omega\rangle$ where $\phi$ represents $g_{\mu\nu}$. And if it does then there would be states of negative norm,  since one can recover the $1/(k^2-M^2)-1/k^2$ propagator by inserting the completeness relation
\begin{eqnarray}
\sum \vert n_1\rangle\langle n_1\vert -\sum \vert n_2\rangle\langle n_2\vert =I
\label{10.1}
\end{eqnarray}
into $i\langle \Omega \vert T[\phi(x)\phi(0)]\vert \Omega\rangle$. However, when Bender and Mannheim  constructed the underlying quantum theory they found that (\ref{10.1}) did not hold. Specifically, they quantized the fourth-order theory and found that the states of the theory were not normalizable. Since the completeness relation given in (\ref{10.1}) assumes normalizable states it is not valid.

With the states not being normalizable the Hamiltonian is not Hermitian (we cannot integrate by parts and throw away surface terms). However, all the poles of the propagator are on the real axis. Thus despite the lack of  Hermiticity all  energy eigenvalues are real. Now Hermiticity is only sufficient for real eigenvalues, with the necessary condition  \cite{bender2010PT,mannheim2018antilinearity} being that the Hamiltonian have an antilinear symmetry such as $PT$ \cite{footnote0}. The theory thus falls into the class of $PT$ theories developed by Carl Bender and collaborators \cite{bender2007making,bender2019pt}. In these theories the bra state is not the Hermitian conjugate of the ket state but the $PT$ conjugate of it. Thus the propagator should be identified as
\begin{eqnarray}
D_{PT}(x)=i\langle \Omega^{PT} \vert T[\phi(x)\phi(0)]\vert \Omega\rangle.
\label{10.2}
\end{eqnarray}
Similarly,  the completeness relation should be given  by
\begin{eqnarray}
\sum \vert n_1\rangle\langle n_1^{PT}\vert +\sum \vert n_2\rangle\langle n_2^{PT}\vert =I,
\label{10.3}
\end{eqnarray}
with all  $\langle n^{PT}\vert m\rangle$ norms being positive definite \cite{bender2008no} even as the insertion of (\ref{10.3}) into (\ref{10.2}) recovers  \cite{bender2008exactly} the $1/(k^2-M^2)-1/k^2$ propagator. 

Since the states are not normalizable we have to continue into the complex plane in order to make them normalizable. Then the theory is well-defined, with the domain of the measure needed for the path integral also having to be continued into the complex plane in order to make it be well-defined too \cite{bender2008exactly,mannheim2018antilinearity}.

\section{One more surprise}

The partial fraction decomposition of the propagator into 
\begin{eqnarray}
\frac{1}{k^4}=\lim_{M^2\rightarrow 0} \frac{1}{M^2}
\left(\frac{1}{k^2-M^2}-\frac{1}{k^2}\right)
\label{11.1}
\end{eqnarray}
becomes singular when we set $M^2=0$. This singular behavior  translates into there not actually being two sets of gravitons after all. The Hamiltonian becomes of nondiagonalizable Jordan-block form and no longer possesses a complete set of eigenstates. Thus we finish  up with only one set of gravitons, and the correct representation of the propagator is as the limit:
\begin{eqnarray}
\frac{1}{(k^2+i\epsilon)^2}=\lim_{M^2\rightarrow 0} \frac{d}{dM^2}
\left(\frac{1}{k^2-M^2+i\epsilon}\right),
\label{11.2}
\end{eqnarray}
with the conformal gravity theory being ghost free and unitary. 

To see what happens to the eigenstates consider $\omega_1=\omega+\epsilon$, $\omega_2=\omega-\epsilon$. In the limit $\epsilon\rightarrow 0$ we obtain 
\begin{eqnarray}
e^{i\omega_1 t}\rightarrow e^{i\omega t},\quad e^{i\omega_2 t}\rightarrow e^{i\omega t}.
\label{11.3}
\end{eqnarray}
Thus both wave functions collapse onto the same wave function, and so we lose an eigenstate. Now this lost eigenstate  has to go somewhere (it cannot simply disappear), so consider 
\begin{eqnarray}
\frac{e^{i\omega_1 t}-e^{i\omega_2 t}}{2\epsilon}\rightarrow it e^{i\omega t}.
\label{11.4}
\end{eqnarray}
This second combination has a well-defined, nonsingular limit, and in the limit becomes nonstationary. So it is no longer an energy eigenstate, and thus we lose an energy eigenstate. 

In regard to the eigenstate that does survive, we note that since the two one-particle states were orthogonal to each other before we took the limit, and since they become the same eigenvector in the limit, in the limit the surviving eigenvector is both parallel to and orthogonal to itself. It  is thus has zero norm.
This is just as we obtained from quantizing conformal gravity, viz.
\begin{eqnarray}
&&[A^{(i)}(\bar{k}),\hat{B}^{(j)}(\bar{k}^{\prime})]=[B^{(i)}(\bar{k}),\hat{A}^{(j)}(\bar{k}^{\prime})]=Z(k)\delta_{i,j}\delta^3(\bar{k}-\bar{k}^{\prime}),
\nonumber\\
&&[A^{(i)}(\bar{k}),\hat{A}^{(j)}(\bar{k}^{\prime})]=0,\quad [B^{(i)}(\bar{k}),\hat{B}^{(j)}(\bar{k}^{\prime})]=0,
\nonumber\\
&& [A^{(i)}(\bar{k}),B^{(j)}(\bar{k}^{\prime})]=0,\quad [\hat{A}^{(i)}(\bar{k}),\hat{B}^{(j)}(\bar{k}^{\prime})]=0,
\label{11.5}
\end{eqnarray}
with the existence of commutators that are zero indicating the presence of zero norm states. Thus the behavior of second-order  plus fourth-order theories is not indicative of the behavior of pure fourth-order theories as the pure fourth-order limit is singular.

\section{Second-order plus fourth-order theories}

\subsection{Setting up the problem}

While our interest in this paper is in the pure fourth-order derivative conformal gravity theory, nonetheless, it is instructive to study second-order plus fourth-order theories. These theories are encountered in quantizing Einstein gravity since graviton loop corrections to the second-order Einstein theory (action linear in the Ricci scalar) generate fourth-order terms (action quadratic in the Ricci scalar or Ricci tensor). To see the issues involved it suffices to consider a second-order plus fourth-order neutral scalar field theory with action and fourth-order derivative equation of motion of the form
\begin{eqnarray}
I_S&=&\tfrac{1}{2}\int d^4x\bigg{[}\partial_{\mu}\partial_{\nu}\phi\partial^{\mu}
\partial^{\nu}\phi-(M_1^2+M_2^2)\partial_{\mu}\phi\partial^{\mu}\phi
+M_1^2M_2^2\phi^2\bigg{]},
\nonumber\\
&&(\partial_t^2-\vec{\nabla}^2+M_1^2)(\partial_t^2-\vec{\nabla}^2+M_2^2)
\phi(x)=0.
\label{12.1}
\end{eqnarray}
The associated propagator obeys
\begin{align}
&(\partial_t^2-\vec{\nabla}^2+M_1^2)(\partial_t^2-\vec{\nabla}^2+M_2^2)D(x)=-\delta^4(x),
\nonumber\\
&D(x)=\int \frac{d^4k}{(2\pi)^4}e^{-ik\cdot x}D(k),
\nonumber\\
&D(k)=-\frac{1}{(k^2-M_1^2)(k^2-M_2^2)}=\frac{1}{(M_1^2-M_2^2)}\left(\frac{1}{k^2-M_2^2}-\frac{1}{k^2-M_1^2}\right).
\label{12.2}
\end{align}
If we identify the $D(k)$ propagator as 
\begin{eqnarray}
D(k)=\int d^4x e^{ik\cdot x}D(x)=i\int d^4x e^{ik\cdot x}\langle \Omega\vert T[\phi(x)\phi(0)]\vert\Omega\rangle,
\label{12.3}
\end{eqnarray}
then insertion of (\ref{10.1}) 
with its ghostlike relative minus sign into $i\langle \Omega\vert T[\phi(x)\phi(0)]\vert \Omega\rangle$ would generate $D(k)$. With the relative minus sign in (\ref{10.1}) there appear to be ghost states in the theory and a  loss of unitarity. 

However, as we had noted above, one cannot simply postulate that  the c-number $D(k)$ is given by the vacuum matrix element  $i\int d^4x e^{ik\cdot x}\langle \Omega\vert T[\phi(x)\phi(0)]\vert\Omega\rangle$ of the quantum field operators. And moreover, one cannot make this postulate just because it is familiar from experience with second-order quantum field theories.  Rather, if it is to be true, then just as in second-order quantum field theory, one needs to quantize the theory, construct the Hilbert space and then take matrix elements. It is precisely this  procedure that has been applied by Bender and Mannheim in \cite{bender2008exactly}. They  constructed the energy-momentum tensor $T_{\mu\nu}$, the canonical momenta $\pi^{\mu}$ and $\pi^{\mu\lambda}$, and the equal-time commutators appropriate to the higher-derivative theory, viz.  
\begin{align}
&T_{\mu\nu}=\pi_{\mu}\phi_{,\nu}+\pi_{\mu}^{\phantom{\mu}\lambda}\phi_{,\nu,\lambda}-\eta_{\mu\nu}{\cal L},
\nonumber\\ 
& \pi^{\mu}=\frac{\partial{\cal L}}{\partial \phi_{,\mu}}-\partial_{\lambda
}\left(\frac{\partial {\cal L}}{\partial\phi_{,\mu,\lambda}}\right)=-\partial_{\lambda}\partial^{\mu}\partial^{\lambda}\phi- (M_1^2+M_2^2)\partial^{\mu}\phi,
\nonumber\\
& \pi^{\mu\lambda}=\frac{\partial {\cal L}}{\partial \phi_{,\mu,\lambda}}=\partial^{\mu}\partial^{\lambda}\phi,
\nonumber\\
&T_{00}=\frac{1}{2}\pi_{00}^2+\pi_{0}\dot{\phi}+\frac{1}{2}(M_1^2+M_2^2)\dot{
\phi}^2-\frac{1}{2}M_1^2M_2^2\phi^2
\nonumber\\
&-\frac{1}{2}\pi_{ij}\pi^{ij}+\frac{1}{2}(
M_1^2+M_2^2)\phi_{,i}\phi^{,i},
\nonumber\\
&\delta(t)[\phi(0),\dot{\phi}(x)]=0, \qquad \delta(t)[\phi(0),\ddot{\phi}(x)]=0, \qquad \delta(t)[\phi(0),\dddot{\phi}(x)]=-i\delta^4(x).
\label{12.4}
\end{align}
With the use of these commutation relations we find that $D(k)=i\int d^4x e^{ik\cdot x}\langle \Omega\vert T[\phi(x)\phi(0)]\vert\Omega\rangle$ is indeed given by (\ref{12.2}), provided of course that $\langle \Omega\vert \Omega\rangle=1$, surprisingly something we will actually have to revisit below. With an eye on what is to follow, we note that we can actually recover (\ref{12.2}) if we set $D(k)=i\int d^4x e^{ik\cdot x}\langle \Omega^{\prime}\vert T[\phi(x)\phi(0)]\vert\Omega\rangle$ with any $\langle \Omega^{\prime}\vert$ serving as the bra as long as it obeys $\langle \Omega^{\prime}\vert\Omega\rangle=1$. However, what we will find is that, innocuous as it may seem, we are not actually free to take the vacuum to be normalizable, and in fact we will find that $\langle \Omega\vert \Omega\rangle$ is infinite. However, the $PT$ vacuum normalization $\langle \Omega^{PT}\vert \Omega\rangle$ will prove to be finite.

We should note that the  identification of $D(k)$  as $D(k)=-1/[(k^2-M_1^2)(k^2-M_2^2)]$ is only formal since $D(k)$ is singular, and to give it a meaning we need to define it via a contour integral and specify the appropriate complex $k_0$ plane contour. If, as is conventional,  we take all of the operators in $H_0=\int d^3x T_{00}$  to be Hermitian, we immediately find that because of the $-(1/2)M_1^2M_2^2\phi^2$ term the Hamiltonian $H_0$ is unbounded from below, the Ostrogradski instability that is characteristic of higher-derivative theories. Now the standard Feynman contour $k^2+i\epsilon$ prescription is chosen so that positive energy states propagate forward in time (viz.  $\omega_1=+(\vec{k}^2+M_1^2)^{1/2}$, $\omega_2=+(\vec{k}^2+M_2^2)^{1/2}$ located below the real $k_0$ axis), while negative energy states propagate backwards ($-\omega_1$, $-\omega_2$ above the real $k_0$ axis). However, having an unbounded from below energy spectrum would entail that  negative energies in one of sectors ($\omega_2$ say) are propagating forward in time (-$\omega_2$ below the real $k_0$ axis), and positive energies are propagating backward in time ($\omega_2$ above the real $k_0$ axis), and would require using an unconventional $i\epsilon$ prescription \cite{bender2008exactly}.  While this possibility is unacceptable physically, it does have the feature that because of the way the singularities are traversed all pole residues are positive. Thus the two options are: bounded energies and negative residues, or unbounded energies and positive residues. These two realizations are inequivalent and correspond to different Hilbert spaces. As discussed in \cite{bender2008exactly} the first option corresponds to working in a Hilbert space in which the relevant $a_2$ effects $a_2\vert \Omega\rangle=0$,  while the latter corresponds to working in a Hilbert space in which $a^{\dagger}_2\vert \Omega\rangle=0$. Thus in no Hilbert space do we have both negative energies and negative residues.

If we wish to use the standard Feynman contour we need to find a way to quantize the theory so that the energy spectrum is bounded from below.  Thus because of the $-(1/2)M_1^2M_2^2\phi^2$  term we cannot take the fields in $H_0$ to be Hermitian. However, all the poles in $D(k)$ occur at real values of $k_0$, and this is precisely the situation that occurs in $PT$ theory, i.e.,  a non-Hermitian Hamiltonian and yet a completely real energy eigenspectrum when the Hamiltonian is $PT$  symmetric. While the explicit treatment is described in the study made in \cite{bender2008no,bender2008exactly}, we note here that this treatment is equivalent to taking $\phi$ to be anti-Hermitian, with the  $-(1/2)M_1^2M_2^2\phi^2$  term then being positive. Technically this is achieved by making a commutation-relation-preserving similarity transform on $\phi(x)$ and equally on $\pi_0$ of the form $S\phi(x)S^{-1}=-i\phi(x)$, $S\pi_0(x)S^{-1}=i\pi_0(x)$, where $S=\exp(\pi\int d^3\bar{x}\pi_0(\bar{x},t)\phi(\bar{x},t)/2)$. While one is always free to make similarity transformations as they preserve eigenvalues and commutation relations, the content of making this particular one is that it transforms the fields into a domain in the complex plane, known as a Stokes wedge, where everything is well behaved, and accordingly there are neither negative energies or negative norms.

\subsection{Quantum-mechanical model}

In \cite{bender2008no,bender2008exactly} both quantum-mechanical and field-theoretic models were discussed. We shall quickly review the  quantum-mechanical case, and then show that its results carry over to quantum field theory. In the quantum-mechanical study given in \cite{bender2008no,bender2008exactly} the spatial dependence of $H_0$ was frozen out by setting $\omega_1=(\vec{k}^2+M_1^2)^{1/2}$, $\omega_2=(\vec{k}^2+M_2^2)^{1/2}$ and fixing $\vec{k}$. Since canonical commutators only involve time derivatives, freezing out the spatial dependence will still give the full dynamical content of the theory. With this freezing out the theory reduces to the two-oscillator Pais-Uhlenbeck (\textrm{PU}) Hamiltonian and canonical commutation relations \cite{bender2008no}

\begin{align}
H_{\rm PU}&=\tfrac{1}{2}p_x^2+p_zx+\tfrac{1}{2}\left(\omega_1^2+\omega_2^2 \right)x^2-\tfrac{1}{2}\omega_1^2\omega_2^2z^2,
\nonumber\\
 [z,p_z]&=i, \qquad [x,p_x]=i,
\label{12.5}
\end{align}
with canonical variables $z$, $p_z$, and $x$, $p_x$. The terms in $H_{\rm PU}$ are in complete parallel to the first four terms in $T_{00}$. And thus we can set $i=[z,p_z]\equiv [\phi,\pi_0]=[\phi,-\dddot{\phi}-(M_1^2+M_2^2)\dot{\phi}]=i\delta^3(x)$, to thus parallel the commutators given in (\ref{12.4}). When the Schr\"odinger problem for $H_{\rm PU}$ was solved in \cite{bender2008no} it was found that the state with energy $(\omega_1+\omega_2)/2$ had a wave function, $\psi_0(z,x)=\exp[\tfrac{1}{2}(\omega_1+\omega_2)\omega_1\omega_2z^2+i\omega_1\omega_2zx-\tfrac{1}{2}(\omega_1+\omega_2)x^2]$, that was not normalizable, with $\langle \Omega\vert \Omega\rangle=\int dzdx\langle \Omega\vert z,x\rangle\langle z,x\vert\Omega\rangle=\int dz dx \psi_0^*(z,x)\psi_0(z,x)$ thus being infinite.  However, it would become normalizable if we replaced $z$ by $-iz$, and thus $p_z$ by $ip_z$. Thus we treat $z$ as being anti-Hermitian rather than Hermitian. This is perfectly legitimate since nothing requires a field to a priori be Hermitian.  Hermiticity is a statement about boundary conditions, with one needing to be able to throw away surface terms in an integration by parts. And while a set of operators might each be Hermitian when acting on their own eigenstates, it does not follow that they would remain Hermitian when acting on the eigenstates of a Hamiltonian that is built out of them. And this is the case when fourth-order derivatives are present, with one not actually knowing before first actually solving the theory and constructing the Hilbert space.

If we replace $z$ by $-iz$ the unbounded from below  $-\tfrac{1}{2}\omega_1^2\omega_2^2z^2$ term in $H_{\rm PU}$ becomes bounded. Thus the Ostrogradski instability only exists if classical fields are real and quantum fields are Hermitian. The replacement of $z$ by $-iz$ thus allows us to use the Feynman contour. On defining $y=SzS^{-1}=-iz$, $q=Sp_zS^{-1}=ip_z$ where $S=\exp(\pi p_zz/2)$, $H_{\rm PU}$ and the commutation relations are transformed into 
\begin{align}
H^{\prime}_{\rm PU}&=\tfrac{1}{2}p_x^2-iqx+\tfrac{1}{2}\left(\omega_1^2+\omega_2^2 \right)x^2+\tfrac{1}{2}\omega_1^2\omega_2^2y^2,
\nonumber\\
 [y,q]&=i, \qquad [x,p_x]=i,
 \label{12.6}
 \end{align}
and now one can take $y$ and $q$ to be Hermitian, with $x$ and $p_x$ remaining Hermitian since they already were (the $\psi_0(z,x)$ wave function is well behaved at large $x$).  At this point the wave functions are well behaved and one can integrate by parts and throw surface terms away. However, this only makes $H^{\prime}_{\rm PU}$ self-adjoint. It is still not Hermitian because of the presence of the factor of $i$ in the $-iqx$ term. While it is not Hermitian the Hamiltonian $H^{\prime}_{\rm PU}$ is $PT$ symmetric. Thus we must use the $PT$ theory inner product, viz. the overlap of a ket with its $PT$ conjugate bra, viz. $\langle \Omega^{PT}\vert=\langle \Omega\vert PC$. Here  $C$ is the  $C$-operator (see e.g. \cite{bender2007making}) that $PT$ theory has, an operator that obeys $[C,H]=0$, $C^2=I$, $[C,PT]=0$.
 And this inner product is positive definite  \cite{bender2008no}, with $\langle \Omega^{PT}\vert\Omega\rangle$ being finite even though $\langle \Omega\vert\Omega\rangle$ is not. With $PT$ symmetry we thus resolve both negative energy and negative norm problems and obtain a fully acceptable, unitary theory.

For our discussion of the scalar field theory we will need to make a 
second quantization of the Pais-Uhlenbeck theory. It is obtained by setting \cite{bender2008exactly}
\begin{align}
&y(t)=-ia_1e^{-i\omega_1t}+a_2e^{-i\omega_2t}-i\hat{a}_1e^{i\omega_1t}+\hat{a}_2
e^{i\omega_2t},
\nonumber\\
&x(t)=-i\omega_1a_1e^{-i\omega_1t}+\omega_2a_2e^{-i\omega_2t}+i\omega_1\hat{a}_1
e^{i\omega_1t}-\omega_2\hat{a}_2e^{i\omega_2t},
\nonumber\\
&p(t)=-\omega_1^2a_1e^{-i\omega_1t}-i\omega_2^2a_2e^{-i\omega_2t}-\omega_1
^2\hat{a}_1e^{i\omega_1t}-i\omega_2^2\hat{a}_2e^{i\omega_2t},
\nonumber\\
&q(t)=\omega_1\omega_2[-\omega_2a_1e^{-i\omega_1t}-i\omega_1a_2e^{-i
\omega_2t}+\omega_2\hat{a}_1e^{i\omega_1t}+i\omega_1\hat{a}_2e^{i\omega_2t}],
\label{12.7}
\end{align}
with the Hamiltonian being given by 
\begin{equation}
H^{\prime}_{\rm PU}=2(\omega_1^2-\omega_2^2)\left(\omega_1^2\hat{a}_1a_1+\omega_2^2
\hat{a}_2a_2\right)+\tfrac{1}{2}(\omega_1+\omega_2),
\label{12.8}
\end{equation}
and the operator commutation algebra being given by
\begin{align}
&[a_1,\hat{a}_1]=[2\omega_1(\omega_1^2-\omega_2^2)]^{-1},\quad
[a_2,\hat{a}_2]=[2\omega_2(\omega_1^2-\omega_2^2)]^{-1},
\nonumber\\
&[a_1,a_2]=0,\quad[a_1,\hat{a}_2]=0,\quad[\hat{a}_1,a_2]=0,\quad
[\hat{a}_1,\hat{a}_2]=0.
\label{12.9}
\end{align}
In (\ref{12.8}) and (\ref{12.9}) the relative signs are all positive (we take $\omega_1>\omega_2>0$ for definitiveness),  so these
equations define a standard positive energy, positive norm, two-dimensional harmonic oscillator system.

\subsection{Quantum field-theoretic model}

Since the scalar field theory also has an Ostrogradski instability, we can anticipate  (and justify below) that for it the  vacuum $\langle \Omega\vert \Omega\rangle$ is infinite. Consequently, we must reinterpret $\phi$ as an anti-Hermitian field (i.e., $\phi$ replaced by $-i\phi$, and accordingly $\pi_0$  replaced by $i\pi_0$), and on doing so we will then give the scalar field theory a bounded energy spectrum. But also the theory will become a $PT$ theory, since while the Hamiltonian might not be Hermitian, all the poles in the $D(k)$ propagator are on the real axis. The theory is thus a $PT$ theory with the Hamiltonian obeying $[H,PT]=0$. In such a case the correct inner product to use (i.e., the one that is time independent) is the overlap of a ket with its $PT$ conjugate bra, viz. $\langle n^{PT}\vert n\rangle=\langle n\vert PC\vert n\rangle$, and unlike the Dirac norm this inner product is positive definite \cite{bender2008exactly}. In addition, the propagator should not be identified with $D(k)=i\langle \Omega\vert T[\phi(x)\phi(0)]\vert \Omega\rangle$  but with  $D_{PT}(x)=i\langle \Omega^{PT}\vert T[\phi(x)\phi(0)]\vert \Omega\rangle$ instead. The completeness relation should not be identified with (\ref{10.1}) but with  (\ref{10.3}), viz. 
\begin{eqnarray}
\sum \vert n_1\rangle\langle n_1^{PT}\vert +\sum \vert n_2\rangle\langle n_2^{PT}\vert =I,
\label{12.10}
\end{eqnarray}
instead. And then the insertion of (\ref{12.10}) into $D_{PT}(x)=i\langle \Omega^{PT}\vert T[\phi(x)\phi(0)]\vert \Omega\rangle$ expressly generates the propagator given in (\ref{12.2}) with its relative minus sign even though there are no states with negative norm \cite{bender2008exactly}.

In terms of creation and annihilation operators, we set $\bar{\phi}=-i\phi$  equal to
\begin{align}
\bar{\phi}(x)=\int \frac{d^3k}{(2\pi)^{3/2}}\left [-ia_{1,\bar{k}}e^{-ik^1\cdot x}+a_{2,\bar{k}}e^{-ik^2\cdot x}-i\hat{a}_{1,\bar{k}}e^{ik^1\cdot x}+\hat{a}_{2,\bar{k}}e^{ik^2\cdot x}\right].
\label{12.11}
\end{align}
The Hamiltonian $H_0$ and commutation relations  are given by \cite{bender2008exactly}
\begin{eqnarray}
\bar{H}&=&\int d^3k\bigg{[}2(M_1^2-M_2^2)(\bar{k}^2+M_1^2)\hat{a}_{1,\bar{k
}}a_{1,\bar{k}}+2(M_1^2-M_2^2)(\bar{k}^2+M_2^2)\hat{a}_{2,\bar{k}}a_{2,
\bar{k}}
\nonumber\\
&&+\frac{1}{2}(\bar{k}^2+M_1^2)^{1/2}+\frac{1}{2}(\bar{k}^2+M_2^2)^{1/2}\bigg{]},
\label{12.12}
\end{eqnarray}
and
\begin{eqnarray}
&& \delta(t)[\dot{\bar{\phi}}(x),\bar{\phi}(0)]=0,\qquad \delta(t)[\ddot{\bar{\phi}}(x),\bar{\phi}(0)]=0,\qquad \delta(t)[\dddot{\bar{\phi}}(x),\bar{\phi}(0)]=-i\delta^4(x),
\nonumber\\
&&[a_{1,\bar{k}},\hat{a}_{1,\bar{k}^{\prime}}]=[2(M_1^2-M_2^2)(\bar{k}^2+
M_1^2)^{1/2}]^{-1}\delta^3(\bar{k}-\bar{k}^{\prime}),\nonumber\\
&&[a_{2,\bar{k}},\hat{a}_{2,\bar{k}^{\prime}}]=[2(M_1^2-M_2^2)(\bar{k}^2+
M_2^2)^{1/2}]^{-1}\delta^3(\bar{k}-\bar{k}^{\prime}),\nonumber\\
&&[a_{1,\bar{k}},a_{2,\bar{k}^{\prime}}]=0,\quad[a_{1,\bar{k}},\hat{a}_{2,
\bar{k}^{\prime}}]=0,\quad[\hat{a}_{1,\bar{k}},a_{2,\bar{k}^{\prime}}]=0,\quad
[\hat{a}_{1,\bar{k}},\hat{a}_{2,\bar{k}^{\prime}}]=0.
\label{12.13}
\end{eqnarray}
With all relative signs being positive (we take $M_1^2>M_2^2$ for definitiveness), there are no states of negative norm or negative energy. 

Given the structure in (\ref{12.12}) and (\ref{12.13}), we see that the limit $M_2^2\rightarrow M_1^2$ is singular, with the theory then becoming Jordan block. The full construction in this limit is given in \cite{bender2008exactly}, with there being only zero norm states and states that are not stationary. These results also carry through to conformal gravity since $W^{\mu\nu}(1)$ as given in  (\ref{6.1})  is diagonal in the metric indices, with (\ref{6.3}) and (\ref{6.4}) exhibiting a Jordan-block structure with zero norm gravitons. 

We can now proceed to study radiative corrections in the scalar field theory. However, first we need to determine whether it  is $\langle \Omega\vert \Omega\rangle$ or $\langle \Omega^{PT}\vert \Omega\rangle$ that is finite.

\section{How to normalize the quantum field theory vacuum}

In treating the simple harmonic oscillator with Hamiltonian $H=\tfrac{1}{2}[p^2+q^2]$ and commutator $[q,p]=i$, one encounters two sets of bases, the wave function basis and the occupation number space basis. The wave function basis is obtained by setting $p=-i\partial/\partial q$ in $H$ and then solving  the Schr\"odinger equation $H\psi(q)=E\psi(q)$. This leads to a ground state with energy $E_0=\tfrac{1}{2}$ and wave function $\psi_0(q)=e^{-q^2/2}$. For occupation number space we set $q=(a+a^{\dagger})/\sqrt{2}$ and $p=i(a^{\dagger}-a)/\sqrt{2}$. This yields $[a,a^{\dagger}]=1$ and $H=a^{\dagger} a+1/2$. We introduce a no-particle state $\vert \Omega \rangle$ that obeys $a\vert \Omega \rangle =0$, with $\vert \Omega \rangle$ being the occupation number space ground state with energy $E_0=\tfrac{1}{2}$.  However, in and of itself this does not fix the norm $\langle \Omega\vert \Omega\rangle$ of the no-particle state or oblige it to be finite.

To fix the norm we need to relate the ground states of the two bases. With $a=(q+ip)/\sqrt{2}$ we set
\begin{align}
\langle q \vert a\vert \Omega \rangle= \frac{1}{\sqrt{2}}\left(q+\frac{\partial}{\partial q}\right)\langle q \vert \Omega \rangle=0,
\label{13.1}
\end{align}
so that $ \langle q \vert \Omega \rangle=e^{-q^2/2}$. We thus identify $\psi_0(q)=\langle q \vert \Omega \rangle$. We now calculate the norm and obtain  
\begin{align}
\langle \Omega \vert \Omega \rangle=\int_{-\infty}^{\infty}dq \langle \Omega \vert q\rangle\langle q\vert \Omega \rangle=
\int_{-\infty}^{\infty}dq \psi^*_0(q)\psi_0(q)
=\int_{-\infty}^{\infty} dq e^{-q^2}=\sqrt{\pi},
\label{13.2}
\end{align}
and thus establish that the norm of the no-particle state is finite. And on setting $\psi_0(q)=e^{-q^2/2}/\pi^{1/4}$ we normalize it to one. That we are able to do this is because we know the form of the wave function $\psi_0(q)$.

However, in the quantum field theory case we do not know the form of the wave function, so we have to proceed differently. We consider the free relativistic scalar field with action 
\begin{eqnarray}
I_{\rm S}=\displaystyle{\int}d^4x \tfrac{1}{2}\left[\partial_{\mu}\phi\partial^{\mu}\phi-m^2\phi^2\right],
\label{13.3}
\end{eqnarray}
with wave equation and Hamiltonian
\begin{eqnarray}
[\partial_{\mu}\partial^{\mu}+m^2]\phi=0,\quad H=\displaystyle{\int } d^3x \tfrac{1}{2}[\dot{\phi}^2+\vec{\nabla}\phi\cdot \vec{\nabla}\phi+m^2\phi^2],
\label{13.4}
\end{eqnarray}
and quantization condition
\begin{align}
[\phi(\bar{x},t),\dot{\phi}(\bar{x}^{\prime},t)]=i\delta^3(\bar{x}-\bar{x}^{\prime}).
\label{13.5}
\end{align}
With $\omega^2_k=\vec{k}^2+m^2$ solutions to the  wave equation  obey 
\begin{eqnarray}
\phi(\vec{x},t)=\int \frac{d^3k}{\sqrt{(2\pi)^3 2\omega_k}}[a(\vec{k})e^{-i\omega_k t+i\vec{k}\cdot\vec{x}}+a^{\dagger}(\vec{k})e^{i\omega_k  t-i\vec{k}\cdot\vec{x}}],
\label{13.6}
\end{eqnarray}
 and on setting  $[a(\vec{k}),a^{\dagger}(\vec{k}^{\prime})]=\delta^3(\vec{k}-\vec{k}^{\prime})$ the Hamiltonian is given by 
 \begin{eqnarray}
 H=\tfrac{1}{2}\int d^3k[\vec{k}^2+m^2]^{1/2} [a^{\dagger}(\vec{k})a(\vec{k})+a(\vec{k})a^{\dagger}(\vec{k})].
 \label{13.7}
 \end{eqnarray}

Given (\ref{13.7}) we can introduce a no-particle state $\vert \Omega\rangle$ that obeys $a(\vec{k})\vert \Omega \rangle=0$ for each $\vec{k}$, and can identify it as the ground state of $H$. However, that does not fix its normalization, and now we cannot realize the canonical commutator as a differential relation, as we cannot satisfy (\ref{13.5}) by setting $\dot{\phi}(\bar{x},t)$ equal to $-i\partial /\partial \phi(\bar{x},t)$ (though we could introduce a functional derivative $\dot{\phi}(\bar{x},t)=-i\delta /\delta \phi(\bar{x},t)$). Thus starting from (\ref{13.5}) we cannot write the Schr\"odinger equation associated with the field-theoretic $H$ as a wave equation, a wave equation that would enable us to determine the normalization of $\vert \Omega\rangle$.

However, there is something that we can do, namely we can reverse engineer what  we did in the quantum-mechanical case.  We thus introduce 
\begin{align}
a(\vec{k})= \frac{1}{\sqrt{2}}[q(\vec{k})+ip(\vec{k})],\quad a^{\dagger}(\vec{k})= \frac{1}{\sqrt{2}}[q(\vec{k})-ip(\vec{k})],
\label{13.8}
\end{align}
so that 
\begin{align}
[q(\vec{k}),p(\vec{k}^{\prime})]=i\delta^3(\vec{k}-\vec{k}^{\prime}),\quad H=\tfrac{1}{2}\int d^3k[\vec{k}^2+m^2]^{1/2} [p^2(\vec{k})+q^2(\vec{k})].
\label{13.9}
\end{align}
We can realize the $[q(\vec{k}),p(\vec{k}^{\prime})]$ commutator by $p(\vec{k}^{\prime})=-i\partial/\partial q(\vec{k}^{\prime})$, and then turn $H$ into a wave operator. In this way  for each $\vec{k}$ we obtain a solution to the Schr\"odinger equation of the form $\psi(\vec{k})=e^{-q^2(\vec{k})/2}/\pi^{1/4}$. We can define a no-particle vacuum for each $\vec{k}$, viz. $\vert \Omega(\vec{k})\rangle$. For each one we have
\begin{align}
\langle q(\vec{k})\vert a(\vec{k})\vert \Omega(\vec{k}) \rangle= \frac{1}{\sqrt{2}}\left[q(\vec{k})+\frac{\partial}{\partial q(\vec{k})}\right]\langle q(\vec{k})\vert \Omega(\vec{k}) \rangle=0,
\label{13.10}
\end{align}
so that $\langle q(\vec{k})\vert \Omega(\vec{k})\rangle=e^{-q^2(\vec{k})/2}/\pi^{1/4}$, and 
\begin{align}
\langle \Omega(\vec{k})\vert \Omega(\vec{k}) \rangle=\int d q(\vec{k})\langle \Omega(\vec{k})\vert q(\vec{k} \rangle\langle q(\vec{k})\vert \Omega(\vec{k}) \rangle
=\int d q(\vec{k})\frac{e^{-q^2(\vec{k})}}{\pi^{1/2}}=1.
\label{13.11}
\end{align}
Then the vacuum for the full $H$ is given by $\vert \Omega\rangle =\Pi_{\vec{k}}\vert\Omega(\vec{k})\rangle$, and it obeys $\langle \Omega\vert\Omega\rangle=1$, to thus have a finite normalization.

The general prescription then is to convert the occupation number space Hamiltonian into a product of individual occupation number spaces each with its own $\vec{k}$, and then determine whether the equivalent wave mechanics ground state wave functions constructed this way have a finite normalization in the conventional Schr\"odinger theory sense. If they do, then so does the full  product vacuum $\vert \Omega \rangle$ of the full $H$. If on the other hand the equivalent wave mechanics wave functions are not normalizable, then neither is the full $\vert \Omega \rangle$. Since the quantum-mechanical Pais-Uhlenbeck ground state $\psi_0(z,x)$ is not normalizable until we transform with the $S=e^{\pi  p_z z/2}$ similarity transformation, the same is true of its relativistic scalar field generalization, to thus justify the need to use its $S=\exp(\pi\int d^3\bar{x}\pi_0(\bar{x},t)\phi(\bar{x},t)/2)$ similarity transformation, to lead us finally to (\ref{12.12}) and (\ref{12.13}), with $\langle \Omega\vert \Omega\rangle$ not being finite, while  $\langle \Omega^{PT}\vert \Omega\rangle$ is.

 \section{Radiative corrections}
 \label{S14}
 \vskip-0.6cm
\begin{figure}[h]
\centering
 \includegraphics[scale=1.2]{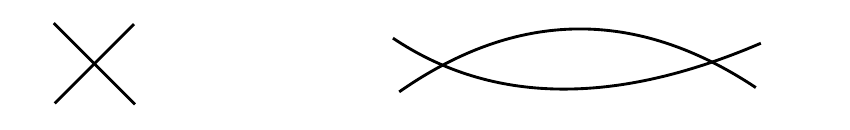}
\caption{Tree Graph and first radiative correction in scalar field theory.} 
\label{loops}
\end{figure}
\vskip-0.6cm
 Having now formulated the scalar field theory,  we can add on a $-\lambda \phi^4$ interaction to $I_S$, to then give a Hamiltonian $H=H_0+\Delta H_0$. In \cite{mannheim2018no} we have given the associated Feynman rules and Landau-Cutkosky cutting rules, and now apply them to the radiative corrections associated with the $-\lambda \phi^4$ interaction.  For radiative corrections we use the propagator given in (\ref{12.2}). Because of its relative minus sign it will lead to Feynman graphs with negative discontinuities and thus threaten unitarity. However,  we have shown that the scalar field $H_0$ Hamiltonian  is free of negative norm states. Then since one cannot change the signature of a Hilbert space in perturbation theory, it must be the case that even with these negative discontinuities the theory remains unitary. We follow \cite{mannheim2018no}, and now explain how this comes about.

We consider the one loop Feynman diagram given in Fig. \ref{loops}. Explicit calculation shows that the total discontinuity across the graph  is given by \cite{mannheim2018no}
 \begin{align}
 \Pi(p_0)=&-\frac{i\lambda^2}{8\pi(M_1^2-M_2^2)^2}\bigg{(}\theta(p_0-2M_2)\frac{(p_0^2-4M_2^2)^{1/2}}{2p_0}
 \nonumber\\
 &+\theta(p_0-2M_1)\frac{(p_0^2-4M_1^2)^{1/2}}{2p_0}
 \nonumber\\
&-\theta(p_0-M_1-M_2)\frac{[p_0^2-(M_1+M_2)^2]^{1/2}[p_0^2-(M_1-M_2)^2]^{1/2}}{p_0^2}\bigg{)},
\label{14.1}
 \end{align}
where the total incoming four-vector is $(p_0,0,0,0)$. As we can see from (\ref{14.1}), the contributions with a positive discontinuity cannot cancel those with a negative discontinuity since the various discontinuities start at different thresholds and have different mass dependences. Nonetheless, cutting the lines in the loop gives no negative norm contributions since $D_{PT}(k)=i\langle \Omega^{PT}\vert T[\phi(x)\phi(0)]\vert \Omega\rangle$ is ghost free.

However, this is not the complete situation, though it would have been had we used the $D(k)=i\langle \Omega\vert T[\phi(x)\phi(0)]\vert \Omega\rangle$ propagator. Specifically, we use the same (\ref{12.2}) propagator regardless of whether we represent it by $D(k)$  or  $D_{PT}(x)$. Thus in the standard approach the negative discontinuity leads to a violation of unitarity.  However, things are different in the $PT$ case. Specifically, we note that the negative discontinuity given in (\ref{14.1}) is of order $\lambda^2$. Thus to cancel it we need some other term of the same order. As noted in \cite{mannheim2018no}, this term is provided by the tree approximation point vertex diagram (the graph shaped like the letter $X$ that also appears in Fig. \ref{loops}), since while it is nominally of order $\lambda$ it is evaluated in states of the form $\langle \Omega^{PT}\vert=\langle\Omega\vert PC$. Now the $C$-operator obeys three conditions: $C^2=I$, $[C,PT]=0$ and $[C,H]=0$.  Of these three conditions the first two are generic. However, the third one depends on the specific Hamiltonian of interest (i.e., the Hamiltonian dynamically determines its own Hilbert space metric and, unlike the Dirac inner product,  it cannot be preassigned). Consequently,  the $C$ required for $[C,H_0+\Delta H_0]=0$ will differ from the $C$ required for $[C,H_0]=0$ by a term of order $\lambda$. Consequently, the order $\lambda$ tree approximation point vertex will acquire a net contribution of order $\lambda^2$. As shown in \cite{mannheim2018no}, when this perturbatively modified tree diagram is taken in conjunction with $\Pi(p_0)$ there is then no net negative discontinuity in the scattering amplitude, and the net discontinuity is unitary.

\section{Other approaches to quantum gravity}

While the focus of this paper is on the pure fourth-order conformal gravity theory, we have also noted that the study given in \cite{bender2008no} shows that second-order plus fourth-order theories also have no ghost states. The much studied $R+R^2$ quantum gravity theory falls into this category, and thus despite its $1/(k^2-M^2)-1/k^2$ propagator it too is ghost free. However, while the relative minus sign in the propagator does not signal a state of negative norm, the propagator still has two poles, a massive pole at $k^2=M^2$ as well as a massless  one at $k^2=0$. Consequently, in addition to the massless graviton there would also be an observable massive particle in the theory. Since no such massive particle has yet been seen, its mass has to be above current detection capabilities, and could be as high as the Planck mass. In contrast, for the pure fourth-order conformal gravity $1/k^4$ propagator there is only one pole (a massless one), with the would-be second one being associated with a nonstationary state.  

Approaches to $R+R^2$ gravity that do not take our $PT$ symmetry analysis into account would use $\langle \Omega \vert$ as the vacuum bra. As well as involve a nonnormalizable vacuum, this approach would involve negative norm states, and then the negative discontinuity associated with the loop graph in Fig. \ref{loops} would not be cancelled. However, making one of the two masses Planck scale would not only explain its current observational absence, it would also postpone the contribution of the negative discontinuity to the same  high scale.  

In this regard it is of interest to recall a study of Hawking and Hertog \cite{hawking2002living}. They made a Euclidean path integral study of the second-plus fourth-order scalar field theory associated with the $I_S$ action given in (\ref{12.1}), and found that while there were both positive and negative norm states there were no transitions between them. Thus, any scattering process with positive norm in-states could only produce positive norm out-states. However, there were still some unitarity violations  in the theory, but such violations could be postponed to a high scale. Their study used a standard real measure.  However, recognizing that we had to continue into the complex plane in order to get normalizable wave functions, we also have to continue the path integral measure into the complex plane. Then one has a well-defined theory with no negative norms or violations of unitarity \cite{bender2008exactly,mannheim2018antilinearity}.

In applications to gravity the $R+R^2$ approaches can be discussed from the perspective of Feynman graviton loop diagrams or path integrals that involve a $D[g_{\mu\nu}]$ measure in which all components of the metric are treated equally. In standard treatments of quantum Einstein gravity one makes a perturbative expansion in the metric and generates multiloop Feynman diagrams. Each perturbative order requires a new counterterm, with the nth-order one being a function of the nth-power of the Riemann tensor and its contractions. With the series not terminating, Einstein gravity is rendered nonrenormalizable.  

An alternate approach has been proposed by 't Hooft \cite{'tHooft2010}. In this approach 't Hooft treated the path integral measure in a nonstandard way, and it enabled him to  find an interesting connection between Einstein gravity and conformal gravity. Specifically, instead of evaluating the path integral as a perturbative series in the metric components $g_{\mu\nu}(x)$, 't Hooft proposes to treat the conformal factor in the metric as an independent degree of freedom. Specifically, he makes a conformal transformation on the (non-conformal-invariant) Einstein-Hilbert action of the form $g_{\mu\nu}(x)= \omega^2(x)\hat{g}_{\mu\nu}(x)$, to obtain 
\begin{equation}
I_{\rm EH}=-\frac{1}{16\pi G}\int d^4x (-\hat{g})^{1/2}\left(\omega^2\hat{R}^{\alpha}_{\phantom{\alpha}\alpha} -6\hat{g}^{\mu\nu}\partial_{\mu}\omega\partial_{\nu}\omega\right),
\label{15.1}
\end{equation}
with everything in $I_{\rm EH}$ now being evaluated in a geometry with metric $\hat{g}_{\mu\nu}(x)$. Then instead of taking the path integral measure to be of the standard $D[g_{\mu\nu}]$ form, 't Hooft proposes that it be taken to be of the form $D[\omega] D[(g_{\mu\nu}/\omega^{2})]=D[\omega] D[\hat{g}_{\mu\nu}]$. 

The utility of this approach is that since the $\omega$ dependence in $I_{\rm EH}$ is quadratic, the $D[\omega]$ path integral can be done analytically. However, in order for the path integral to be bounded one needs to give $\omega$ an imaginary part \cite{footnote1}.  With this choice, the $\omega$ path integral will generate an effective action $I_{\rm EFF}$ of the form ${\rm Tr~ln}[\hat{g}^{\mu\nu}\hat{\nabla}_{\mu}\hat{\nabla}_{\nu}+(1/6)\hat{R}^{\alpha}_{\phantom{\alpha}\alpha}]$, and after dimensional regularization  is found not to generate an infinite set of divergent terms at all, but rather to only generate just one divergent term, viz. the logarithmically divergent
\begin{equation}
I_{\rm EFF}=C\int d^4x (-\hat{g})^{1/2}[\hat{R}^{\mu\nu}\hat{R}_{\mu\nu}-{1 \over 3}(\hat{R}^{\alpha}_{\phantom{\alpha}\alpha})^2],
\label{15.2}
\end{equation}
with $C$ being the very same logarithmically divergent constant that, other than (an undisplayed)  kinematic factor, had appeared in (\ref{8.3}), to which (\ref{15.2}) is  otherwise identical.

In (\ref{15.2}) we immediately recognize the conformal gravity action. Since the action in (\ref{15.1}) is the same action as that obeyed by a conformally coupled scalar field, and the $\omega$ path integral measure is the same as that of a scalar field, everything is conformal, and the $\omega$ path integral can only generate a conformal invariant effective action -- hence (\ref{15.2}).  Through the unusual treatment of the conformal factor we thus find a connection between Einstein gravity and conformal gravity. 

From the perspective of Einstein gravity, the utility of (\ref{15.2}) is that since conformal gravity is renormalizable, the subsequent $D[\hat{g}_{\mu\nu}]$ integration of $I_{\rm EFF}$ should not generate any additional counterterms, while the nonleading terms contained in ${\rm Tr~ln}[\hat{g}^{\mu\nu}\hat{\nabla}_{\mu}\hat{\nabla}_{\nu}+(1/6)\hat{R}^{\alpha}_{\phantom{\alpha}\alpha}]$ would generate contributions to the path integration that could potentially be finite. One still has to deal with the divergent $C$ term in (\ref{15.2}), and rather than have it renormalized (say by adding on an intrinsic conformal term of the form given in (\ref{5.2})), 't Hooft explores the possibility that $C$ remain uncanceled, and we refer the reader to \cite{'tHooft2010} for details. (With $C$ appearing with the same overall sign in (\ref{8.3}) and (\ref{15.2}), and with a gauge or scalar field path integration over their kinetic energy actions having the same sign too \cite{'tHooft2010}, an interplay between fermionic and bosonic fields cannot cancel $C$ -- with massless superpartner fields not being able to effect the same cancellation in this particular  curved space background that they can achieve in a flat one.)

Apart from these $R+R^2$ type studies there are quantum gravity studies that do not consider fourth-order derivative terms at all, the most prominent being string theory and loop quantum gravity. However, as the fermion Dirac action path integration $\int D[\psi]D[\bar{\psi}]\exp{iI_{\rm D}}$ leads to the fourth-order action given in (\ref{8.3}), in the end  fourth-order issues cannot be avoided.

 All of  these approaches to quantum gravity always involve the Einstein-Hilbert action. In contrast, the conformal gravity theory studied here does not involve the Einstein-Hilbert action at all, as it is a pure fourth-order derivative approach.   And in a sense it is in this regard that it most differs from all other approaches to the quantum gravity problem.

 \section{Summary}

The Einstein equations were first derived for solar system physics. If we continue the Einstein equations to galaxies we get the dark matter problem. If we continue the Einstein equations to cosmology  we get the cosmological constant problem. If we quantize the Einstein equations we get renormalizability  and zero-point problems.
Conclusion: the standard extrapolation of solar system wisdom is not reliable. 

If all these problems thus have a common origin, then maybe they have a common solution. So we seek an alternate extrapolation. Hence conformal gravity, a theory that contains the Schwarzschild solution and controllable departures from it, that is renormalizable and  unitary, and that forbids any fundamental cosmological constant at the level of the Lagrangian.
 
In a renormalizable theory of gravity the equations of motion can be treated as well-defined operator identities of the form:
\begin{eqnarray}
(T^{\mu\nu}_{\rm GRAV})_{\rm DIV}+(T^{\mu\nu}_{\rm M})_{\rm DIV}=0,\qquad
(T^{\mu\nu}_{\rm GRAV})_{\rm FIN}+(T^{\mu\nu}_{\rm M})_{\rm FIN}=0.
\label{16.1}
\end{eqnarray}
They couple the gravity sector infinities to those in the matter sector, with the matter sector quantization forcing the gravity sector to be quantized so as to cause all these infinities to mutually cancel each other while fixing the gravitational renormalization constant $Z$ in the gravity sector.
These equations hold in no matter what states one takes matrix elements, but in a mass generating spontaneously-broken vacuum the gravitational $Z$ will readjust  \cite{mannheim2017mass} so as to then control the cosmological constant that is induced by the breaking.

If all mass and length scales come from symmetry breaking, then all scales come from quantum mechanics, i.e., without length scales there cannot be any curvature. Thus spacetime curvature is intrinsically quantum-mechanical and gravity is quantized by its matter source.

In order to construct a renormalizable and unitary theory of quantum gravity we have used an approach based on $PT$ symmetry. The essence of this approach can be categorized by just one key ingredient, namely in its Hilbert space inner product. Nothing in physics requires that one must use the Dirac inner product (the overlap of a ket with its Hermitian conjugate bra). For a theory all of whose energy eigenvalues are real and all of whose energy eigenvectors are complete, any inner product that is finite, time independent and positive will suffice. However, what distinguishes the $PT$ theory  inner product (the overlap of a ket with its $PT$ conjugate bra) from the Dirac inner product is that while the Dirac inner product can be preassigned in a  Hamiltonian-independent way, the $PT$ inner product $\langle \psi\vert PC\vert\psi\rangle$ is determined by the theory itself, with  the $C$-operator having to depend each time on the Hamiltonian of interest, since each time the appropriate $C$-operator is required to commute with the relevant $H$. Thus the Hamiltonian fixes its own Hilbert space, and the choice is not discretionary. In this sense it parallels General Relativity, since the spacetime metric is also not preassigned but is fixed each time by the solution to the relevant gravitational equations of motion. 

With the $PT$ inner product being dependent on the Hamiltonian of interest,  as one adds a perturbation the inner product readjusts. It is because of this that radiative corrections to second-order plus fourth-order theories maintain unitarity if one works in a Hilbert space with the $PT$ theory inner product.

Moreover, there is an additional aspect to $PT$ symmetry, one that cannot be obtained with Hermitian Hamiltonians at all, namely one can have Jordan-block Hamiltonians that cannot be diagonalized at all. The pure fourth-order derivative conformal gravity theory falls into this category, and it is this aspect of it that enables the theory to be a consistent and unitary theory of quantum gravity.

To conclude, we note that at the beginning of the 20th century studies of black-body radiation on microscopic scales led to a paradigm shift in physics. Thus it could be that at the beginning of the 21st century studies of black-body radiation, this time on macroscopic cosmological scales, might be presaging a paradigm shift all over again.

\smallskip

\textbf{Data availability}
Data sharing not applicable to this article as no datasets were generated or analyzed during the current study.


\begin{thebibliography}{99}


\bibitem{bender2008no} Bender, C.M., Mannheim, P.D.:  No-ghost theorem for the fourth-order derivative Pais-Uhlenbeck oscillator model. Phys. Rev. Lett. \textbf{100}, 110402 (2008) 
\url{https://doi.org/10.1103/PhysRevLett.100.110402}


\bibitem{bender2008exactly} Bender, C.M., Mannheim, P.D.:  Exactly solvable $PT$-symmetric Hamiltonian having no Hermitian counterpart. Phys. Rev. D \textbf{78} 025022 (2008). 
\url{https://doi.org/10.1103/PhysRevD.78.025022}


\bibitem{colella1975observation} Colella, R. Overhauser, A.W., Werner, S.A.: Observation of gravitationally induced quantum interference. Phys. Rev. Lett. \textbf{34}, 1472 (1975). 
\url{https://doi.org/10.1103/PhysRevLett.34.1472}

\bibitem{mannheim1998classical} Mannheim, P.D.: Classical underpinnings of gravitationally induced quantum interference. Phys. Rev. A \textbf{57},  1260 (1998). 
\url{https://doi.org/10.1103/PhysRevA.57.1260}



\bibitem{mannheim2017mass} Mannheim, P.D.: Mass generation, the cosmological constant problem, conformal symmetry, and the Higgs boson. Prog. Part. Nucl. Phys. \textbf{94}, 125 (2017). 
\url{https://doi.org/10.1016/j.ppnp.2017.02.001}



\bibitem{weinberg1972gravitation} Weinberg, S.: Gravitation and Cosmology:
Principles  and Applications of the General Theory of Relativity. Wiley, New York (1972)

\bibitem{mannheim2006gauge} Mannheim, P.D.: Gauge invariant treatment of the energy carried by a gravitational wave. Phys. Rev. D \textbf{74}, 024019 (2006). 
\url{https://doi.org/10.1103/PhysRevD.74.024019}{2006}

\bibitem{nambu1961dynamical} Nambu, Y., Jona-Lasinio, G.: Dynamical model of elementary particles based on an analogy with superconductivity. I. Phys. Rev. \textbf{122}, 345 (1961). \url{https://doi.org/10.1103/PhysRev.122.345}

\bibitem{mannheim1989exact} Mannheim, P.D., Kazanas, D.:  Exact vacuum Solution to conformal Weyl gravity and galactic rotation curves. Astrophys. J. 342, 635 (1989). \url{https://doi.org/10.1086/167623}

\bibitem{mannheim2006alternatives} Mannheim, P.D.: Alternatives to dark matter and dark energy. Prog. Part. Nucl. Phys. \textbf{56}, 340 (2006). \url{https://doi.org/10.1016/j.ppnp.2005.08.001}

 \bibitem{mannheim2011imp} Mannheim, P.D., O'Brien. J.G.: Impact of a global quadratic potential on galactic rotation curves. Phys. Rev. Lett. \textbf{106}, 121101 (2011). \url{https://doi.org/10.1103/PhysRevLett.106.121101}


 \bibitem{mannheim2012fitgal} Mannheim, P.D., O'Brien. J.G.: Fitting galactic rotation curves with conformal gravity and a global quadratic potential. Phys. Rev. D \textbf{85}, 124020 (2012). \url{https://doi.org/10.1103/PhysRevD.85.124020}

 
\bibitem{obrien2012fitdwa} O'Brien, J.G., Mannheim, P.D.,: Fitting dwarf galaxy rotation curves with conformal gravity.  Mon. Not. R. Astron. Soc. \textbf{421,} 1273 (2012). \url{https://doi.org/10.1111/j.1365-2966.2011.20386.x}




\bibitem{mannheim2011comprehensive} Mannheim, P.D.: Comprehensive solution to the cosmological constant, zero-point energy, and quantum gravity problems. Gen. Rel. Gravit. \textbf{43}, 703 (2011).  
\url{https://doi.org/10.1007/s10714-010-1088-z}


\bibitem{Wilczek1982families} Wilczek, F., Zee, A.: Families from spinors. Phys. Rev. D \textbf{25}, 553 (1982). 
\url{https://doi.org/10.1103/PhysRevD.25.553}

\bibitem{Riess1998observational} Riess, A.G.  et. al.: Observational evidence from supernovae for an accelerating universe and a cosmological constant. Astronom. J. \textbf{116}, 1009 (1998). 
\url{https://doi.org/10.1086/300499}

\bibitem{Perlmutter1998measurements} Perlmutter, S.  et. al.: Measurements of $\Omega$ and $\Lambda$  from 42 high-redshift supernovae.  Astrophys. J. \textbf{517}, 565 (1999). 
\url{https://doi.org/10.1086/307221}



\bibitem{hooft2015local} 't Hooft, G.: Local conformal symmetry: The missing symmetry component for space and time. Int. J. Mod. Phys. D   \textbf{24}, 1543001 (2015). 
\url{https://doi.org/10.1142/S0218271815430014} 

\bibitem{'tHooft2010} 't Hooft,  G.:  Probing the small distance structure of canonical quantum gravity using the conformal group, arXiv:1009.0669 [gr-qc], \url{https://doi.org/10.48550/arXiv.1009.0669}; The conformal constraint in canonical quantum gravity, arXiv:1011.0061 [gr-qc], \url{https://doi.org/10.48550/arXiv.1011.0061};  A class of elementary particle models without any adjustable real parameters, Found. Phys. \textbf{41}, 1829 (2011).   \url{https://doi.org/10.1007/s10701-011-9586-8} 


 \bibitem{bender2010PT} Bender, C.M., Mannheim, P.D.:  $PT$ symmetry and necessary and sufficient conditions for the reality of energy eigenvalues, Phys. Lett. A \textbf{374}, 1616 (2010).  
\url{https://doi.org/10.1016/j.physleta.2010.02.032}
 
 \bibitem{mannheim2018antilinearity} Mannheim, P.D.: Antilinearity rather than Hermiticity as a guiding principle for quantum theory. 
J. Phys. A: Math. Theor. \textbf{51}, 315302 (2018). \url{http://iopscience.iop.org/article/10.1088/1751-8121/aac035/meta}

\bibitem{footnote0}  If $AHA^{-1}=H$, where $A$ is an antilinear operator such as $PT$ and  $H\vert \psi\rangle=E\vert \psi\rangle$, then 
 $AH\vert \psi\rangle=AE\vert \psi\rangle=E^*A\vert \psi\rangle=AHA^{-1}A\vert \psi\rangle=HA\vert \psi\rangle$. Thus for every eigenvalue $E$ there is an eigenvalue $E^*$, with all the poles of $D(k)$ or $D_{PT}(k)$ being  in the $E=E^*$ realization in which  all energies are real.



 \bibitem{bender2007making} Bender, C.M.:  Making sense of non-Hermitian Hamiltonians. Rep. Prog. Phys. \textbf{70} 947 (2007).  
 \url{https://doi.org/10.1088/0034-4885/70/6/R03}
 
 \bibitem{bender2019pt} Bender, C.M.:  $PT$ Symmetry in Quantum And Classical Physics. World Scientific, Singapore (2019).
 
 \bibitem{mannheim2018no} Mannheim, P.D.: Unitarity of loop diagrams for the ghostlike $1/(k^2-M_1^2)-1/(k^2-M_2^2)$ propagator, Phys. Rev. D \textbf{98}, 045014 (2018). 
\url{https://doi.org/10.1103/PhysRevD.98.045014}

\bibitem{hawking2002living} Hawking, S.W. and Hertog, T.: Living with ghosts, Phys. Rev. D \textbf{65}, 103515 (2002). \url{https://doi.org/10.1103/PhysRevD.65.103515}
 
 
 
 \bibitem{footnote1} This is similar in spirit to the $PT$  approach, where the path integral measure is also continued into the complex plane. While study of it is beyond the scope of this paper,  the approach of 't Hooft might fall into the $PT$ category.
 
 
\end{thebibliography}
\end{document}